\documentclass[superscriptaddress,aps,pra,twocolumn,showpacs,nofootinbib,longbibliography, showkeys]{revtex4-2}
\usepackage{amsmath,amssymb,amsthm}
\usepackage{easybmat}
\usepackage[colorlinks=true,citecolor=blue,urlcolor=blue, linkcolor= magenta]{hyperref}
\usepackage[pdftex]{graphicx}
\usepackage{times,txfonts}
\usepackage{braket}
\usepackage{color}
\usepackage{natbib}
\newcommand{\be}{\begin{equation}}
	\newcommand{\ee}{\end{equation}}
\newcommand{\ba}{\begin{eqnarray}}
	\newcommand{\ea}{\end{eqnarray}}

\begin{document}
	
	%\preprint{APS/123-QED}
	\title{Quantum chaos in the Dicke model and its variants}
	
\author{Devvrat Tiwari\textsuperscript{}}
	\email{devvrat.1@iitj.ac.in}%Lines break automatically or can be forced with \\
\author{Subhashish Banerjee\textsuperscript{}}
	\email{subhashish@iitj.ac.in }
	\affiliation{Indian Institute of Technology Jodhpur-342030, India\textsuperscript{}}

\date{\today}

\begin{abstract}
Recently, the out-of-time-ordered correlator (OTOC) has gained much attention as an indicator of quantum chaos. In the semi-classical limit, its exponential growth rate resembles the classical Lyapunov exponent. The quantum-classical correspondence has been supported for the one-body chaotic systems as well as realistic systems with interactions, as in the Dicke model, a model of multi-two-level atoms and cavity field interactions. To this end, we calculate the OTOC for different variations of the Dicke model in an open quantum system setting. The connection between the superradiant phase transition of the Dicke model and the OTOC is studied. Further, we establish a relation between the OTOC and the second-order coherence function. This becomes important for the experimental studies of the OTOC and quantum chaos in the models of quantum optics. 
\end{abstract}

%Use showkeys class option if keyword
%display desired
\maketitle

%\tableofcontents
\section{Introduction}\label{sec-intro}
The study of quantum chaos seeks to find a relationship between the classical and quantum dynamics of chaotic systems~\cite{casati1979stochastic, SHEPELYANSKY1983208, haake2001quantum, Toda_1988, gutzwiller1991chaos, Zurek_1994, Rigol_2016, Christensen_1998}.  In the limit of large quantum numbers, classical mechanics is shown by the correspondence principle as the classical limit of quantum mechanics. Therefore, classical chaos should also have an underlying quantum mechanism. Quantum chaos seeks answers to this question of the relationship between quantum mechanics and classical chaos~\cite{gutzwiller1991chaos, haake2001quantum}. Random matrix theory ideas~\cite{mehta2004random, Pandey_1981, porter1965statistical}, formulated by Wigner~\cite{Wigner_1967} and Dyson~\cite{Dyson_1962} to understand the spectra of complex atomic nuclei, were used traditionally to probe quantum systems with chaotic classical counterparts~\cite{Bohigas_book, haake2001quantum, Soshnikov_2001, pandey2019quantum}.  Various strategies have been created to bridge the gap between classical and quantum depictions of chaos by using features that are similar in both realms. This includes techniques, such as the level statistics of the quantum states~\cite{Bohigas_1984, Ullmo_2016}, the structure of the eigenstates~\cite{CHIRIKOV198568, FLAMBAUM1999205}, the exponential increase of complexity~\cite{Peres_1996, Izrailev_2019}, and the exponential decay of the Loschmidt echo (a measure of the revival that occurs when an imperfect time-reversal procedure is applied to a complex quantum system)~\cite{Jalabert_2001, GORIN2006, Vallejos_2002}.

An idea of quantum information scrambling (a process of spreading and effective loss of quantum information in a quantum many-body system) by black holes was brought out~\cite{Patrick_Hayden_2007, Yasuhiro_Sekino_2008}, which was popularized and given a great fillip~\cite{Shenker2014}. This scrambling can be measured using out-of-time-ordered correlators (OTOC), which was conjectured to have a bound on its growth rate~\cite{Maldacena2016, Maldacena2, Xu2020}. The OTOC, first established in the context of superconductivity~\cite{otoc_Ovchinnikov}, is now presented as a measure of quantum chaos, with its growth rate attributed to the classical Lyapunov exponent~\cite{Hashimoto2017, Rozenbaum_2017, ueda1, ignacio, ueda_3, Felix_2019}. Recently, OTOC has started receiving considerable attention because of its appearance as a valuable tool for generalizing quantum chaos studies beyond the time-independent, one-body case. Not only that, it has been extensively studied in a wide variety of systems, including quantum thermalization~\cite{Srednicki1, Srednicki2}, Sachdev-Ye-Kitaev model~\cite{Kitaev1, Sachdev_1993, Maldacena2, Zhang2021Quantum}, quantum field theories~\cite{stanford_2015, Stanford2016, chowdhury_2017, Swingle_2017}, random unitary models~\cite{Nahum_2018, khemani_2018, Rakovszky, weinstein2022}, and spin chains~\cite{Luitz2017, Fortes_2020, Heyl, lin_2018}. Moreover, the OTOC has also been measured experimentally using techniques of nuclear magnetic resonance~\cite{Garttner2017-bs, Li_2017, Wei_2018, Niknam_2020}.

Atom-cavity field interactions are critical to science and technology and provide a foundation for the field of quantum optics~\cite{loudon, mandel-wolf, scully-zubairy, gs-agarwal, Larson}. The first quantum mechanical model describing the interaction between a single atom and a cavity of electromagnetic field was given by Jaynes and Cummings~\cite{jaynes-cummings, shore-knight}. Here, we discuss a class of quantum optical models, which are multi-atom generalizations of the Jaynes-Cummings model. This includes variations of the model proposed by Dicke in 1954~\cite{Dicke1} to describe super-radiance, a field-mediated collective emission of an ensemble of excited atoms or ions~\cite{HEPP-Leib, Hepp_Leib2, wang-hioe, Hioe2, CARMICHAEL_Gardiner, Garraway}. The superradiant phase transition of the Dicke model has been studied in recent experiments~\cite{Baumann2010, Dimer_2007, Zhiqiang_17, Domokos, Baden_2014}. The presence of quantum chaos and quantum phase transition in the Dicke model has long been an area of interest~\cite{emary1, emary2, Haake2, alavirad, bastarrachea2015chaos, Hirsch_2016, Hirsch_2022, Wang_2022}. A demonstration of quantum-classical correspondence, using OTOC for the Dicke model, was made in the unitary case~\cite{Hirsch_2019}. Further, the Dicke model under dissipation was studied to explore the connections between the dissipative phase transition and non-Hermitian random matrix theory in~\cite{mahaveer_chaos}. Here, we endeavor to explore the chaotic behavior of the variations of the Dicke model in an open quantum system setting with dissipation. 

A realistic quantum system interacts with its surroundings. These interactions significantly change the dynamics of the quantum system. There is a loss of information from the system to the environment involved. A framework to study such systems is provided by the theory of open quantum systems (OQS)~\cite{breuer-petruccione, sbbook, weiss}. The ideas of open quantum systems find use in a wide variety of applications~\cite{Louisell, caldeira-leggett1983, GrabertSchrammIngold, sbqbm, sbsterngerlach, sbrichard, sbjavidprd, sbkhushboocoherence, sbrichard, sbjavidprd, sbkhushboocoherence, Bhattacharya2018, Utagi2020, Kumar_2018, reactioncoordinaterefs1, reactioncoordinaterefs2, reactioncoordinaterefs3, blhu, sgad, plenio}. Exploring quantum chaos and OTOC in open quantum systems is an ongoing area of research. Several techniques have been developed to address this issue~\cite{Zhang_2019, Tuziemski_2019, Deffner_2021, Zanardi_2021, Han_2022, Zanardi_2023}. To this end, we use the adjoint form of the GKSL master equation~\cite{breuer-petruccione, sbbook} to study the operator evolution in time, which is then used to calculate the OTOC. To calculate the OTOC in Markovian open quantum systems, an extension of the quantum regression theorem was derived in~\cite{Molmer_2019}, where an indirect connection between the second-order coherence $g^{(2)}(t)$ function and OTOC was given. To this effect, we try to establish a direct relationship between the characteristics of quantum optics ($g^{(2)}(t)$ function) and the chaotic behavior of the quantum optical models. 

The paper is structured as follows: in Sec. \ref{sec-prelim}, we discuss the preliminaries of the quantum optical models, such as their quantum phase transitions and the out-of-time-ordered correlator (OTOC) and its various forms. Section \ref{sec-otoc_behavior} discusses the behavior of the OTOC in the models discussed here with the calculation of the Lyapunov exponents and the effect of quantum phase transition on the chaotic behavior of these models. The behavior of OTOC in short, intermediate, and long-time is also discussed. In Sec. \ref{sec-otoc_g2}, we highlight the connection between the $g^{(2)}(t)$ function and the OTOC and discuss the characteristics of light in conjunction with quantum chaos. This is followed by the Conclusions.

\section{Preliminaries}\label{sec-prelim}
In this section, we briefly discuss the many-body quantum optical models, particularly the models where $N$-qubits interact with QED-cavity. In order to understand the chaotic properties of these models, we will make use of the out-of-time-ordered-correlators, which are briefly described here.  

\subsection{Quantum optical models}\label{sec-quantum_opt_models}
Here, we discuss the various forms of the multi-atom-field interaction model, i.e., the Dicke model. In particular, we study the generalized Dicke model, the $N$-qubit Dicke model, the Tavis-Cummings model, and the Floquet Dicke model. 

\subsubsection{The generalized Dicke (GD) model}\label{sec-gen_Dicke_model}
The generalized Dicke (GD) model consists of $N$ two-level atoms with transition frequencies $\omega_a$, coupled to a single mode of a quantized radiation field of frequency $\omega_c$. The Hamiltonian (for $\hbar = 1$) of this model is
\begin{align}
    H_{GD} &= \omega_c \hat{a}^\dagger\hat{a} + \frac{\omega_a}{2} \sum_{j = 1}^N \sigma_j^z + \frac{\lambda}{\sqrt{N}}\sum_{j = 1}^N\left(\hat{a}\sigma^+_j + \hat{a}^\dagger \sigma^-_j\right) \nonumber \\ 
    &+ \frac{\lambda'}{\sqrt{N}}\sum_{j = 1}^N\left(\hat{a}\sigma^-_j + \hat{a}^\dagger \sigma^+_j\right), 
    \label{eq-gen_dicke_Ham_all}
\end{align}
which, upon using the collective angular momentum operators $J_i = \frac{1}{2}\sum_{j=1}^N\sigma_j^i$ ($\sigma^i$ are the Pauli matrices with $i = {x, y, z}$) and $J_{\pm} = \sum_{j=1}^N\sigma^{\pm}$, is given as 
\begin{align}
    H_{GD} &= \omega_c \hat{a}^\dagger\hat{a} + \omega_a J_z + \frac{\lambda}{\sqrt{N}}\left(J_+\hat{a} + J_-\hat{a}^\dagger \right) + \frac{\lambda'}{\sqrt{N}}\left(J_-\hat{a} + J_+\hat{a}^\dagger\right). 
    \label{eq-gen_dicke_Ham}
\end{align}
The collective angular momentum operators $J_i$ ($i=z,+,-$), in terms of a pseudospin of length $j=N/2$, describe the ensemble of two-level atoms. The single mode quantized radiation field is depicted by the number operator $\hat{a}^\dagger \hat{a}$, where $\hat{a}$, and $\hat{a}^\dagger$ are the bosonic creation and annihilation operators. The third and fourth terms depict the coupling between the quantized radiation field and the atoms for co- and counter-rotating terms, respectively. Here $\lambda$ denotes the interaction strength for the co-rotating term, and $\lambda'$ denotes the interaction strength for the counter-rotating term. $\sqrt{N}$ is the scaling factor.

\begin{figure}[!htb]
    \centering
    \includegraphics[width = 1\columnwidth]{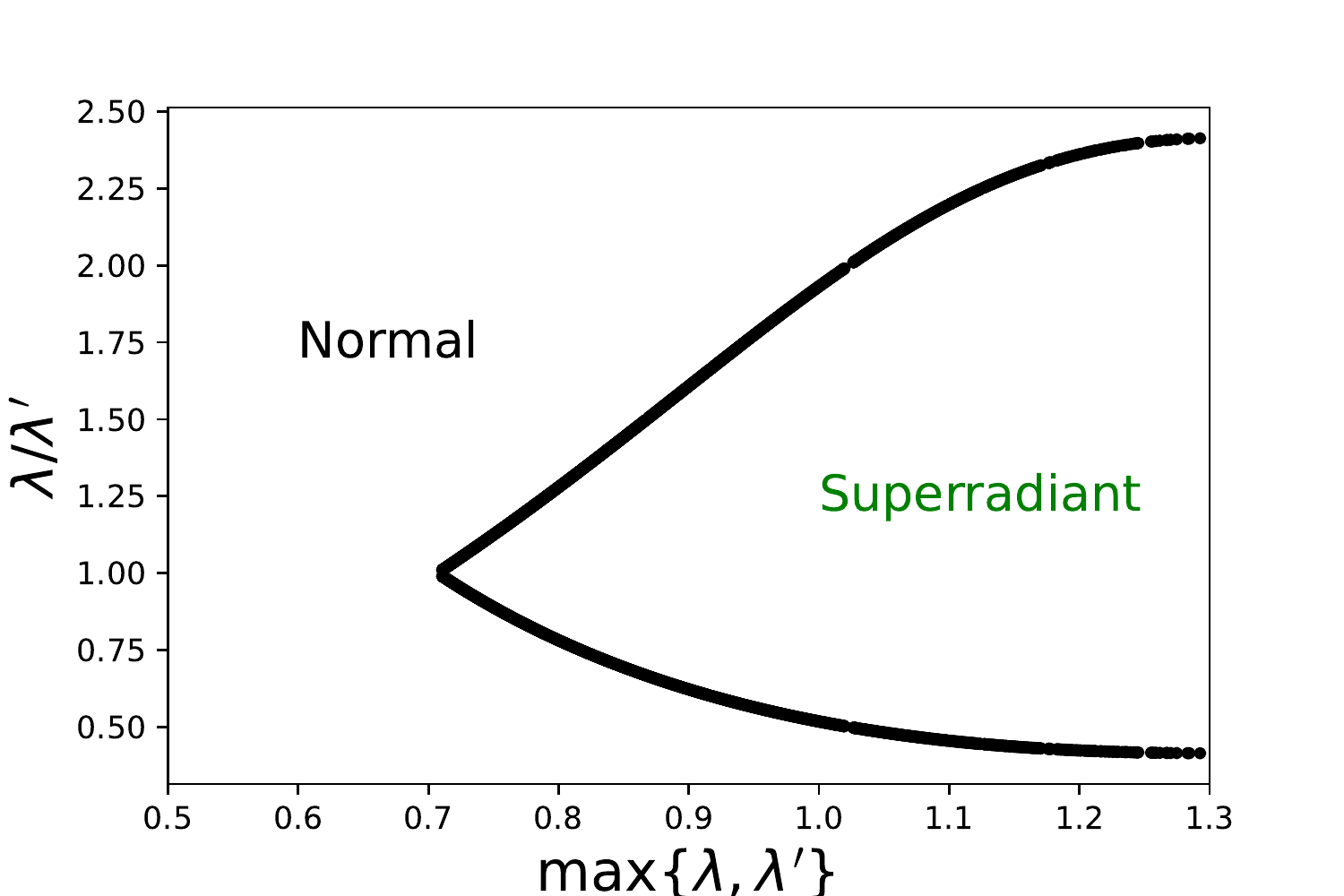}
    \caption{Phase transition in the generalized Dicke (GD) model. The region marked with green color shows the superradiant phase, and the one marked with black color shows the normal phase. The parameters are $N = 7$, $\omega_a = \omega_c = 1$, $\kappa = 1,~\gamma = 0,~s_z = -1/2$. This has been derived from~\cite{Demler}.}
    \label{Gen_Dicke_model_Phase_transtion}
\end{figure}

Note that if we keep the interaction strengths $\lambda'$ equal to $\lambda$, the above model boils down to the Dicke model~\cite{Dicke1}, and for $\lambda' = 0 $, it becomes the Tavis-Cummings (TC) model~\cite{tavis-cummings, tavis-cummings2}. These models are discussed below. The phase transition of the generalized Dicke model is well studied in the literature~\cite{Hioe2, Demler, Zhiqiang_17, Kirton_2019}. The superradiant phase transition is one of the prominent features of the Dicke model. For the generalized Dicke model with dissipation, the phase transition manifests itself as a dynamical instability of the cavity and, hence, can be detected by considering its response function, in particular the retarded Green’s function~\cite{Demler}. At the superradiant phase transition, the response becomes an exponentially growing function of time, indicating a dynamical instability. Mathematically, this corresponds to the poles of the Green’s function, which in turn correspond to the zeros of the inverse Green’s function.
This leads to the critical point given by 
\begin{align}
\label{eq-critical_phase_trans}
    \lambda_c^2 =& \frac{(R-1)(\omega_a\omega_c + R\kappa\gamma)}{4R^2s_z}\left(1 - \sqrt{1 - \frac{R^2(\omega_c^2+\kappa^2)(\gamma^2 + \omega_a^2)}{(\omega_c\omega_a+R\kappa\gamma)^2}}\right),
\end{align}
for a fixed $\lambda/\lambda'$ ratio, where $R = \frac{1 - (\lambda'/\lambda)^2}{1 + (\lambda'/\lambda)^2}$, $\gamma$ is the decay rate of the atomic system, and $\kappa$ is the decay rate of the cavity. The value of $s_z = \langle\sigma_z\rangle$ corresponds to the initial state, which is $-1/2$ for a fully polarized state. In Fig.~\ref{Gen_Dicke_model_Phase_transtion}, we can see the phase transition of the generalized Dicke model between the normal and the superradiant phases. It can be observed that at $\lambda = \lambda'$, the minimum critical coupling is achieved. This corresponds to the case of the $N$-qubit Dicke model discussed below. The critical coupling is larger for the case when either of the counter- or co-rotating terms dominate. In the case, when $\lambda\gg \lambda'$, the model becomes equivalent to the TC model, and the super-radiance can not be achieved (under dissipation) as discussed below in Sec. \ref{sec-TC_model}.
\subsubsection{The $N$-qubit Dicke model}\label{sec-N_qubit_Dicke_model}
\begin{figure}[!htb]
    \centering
    \includegraphics[width = 1\columnwidth]{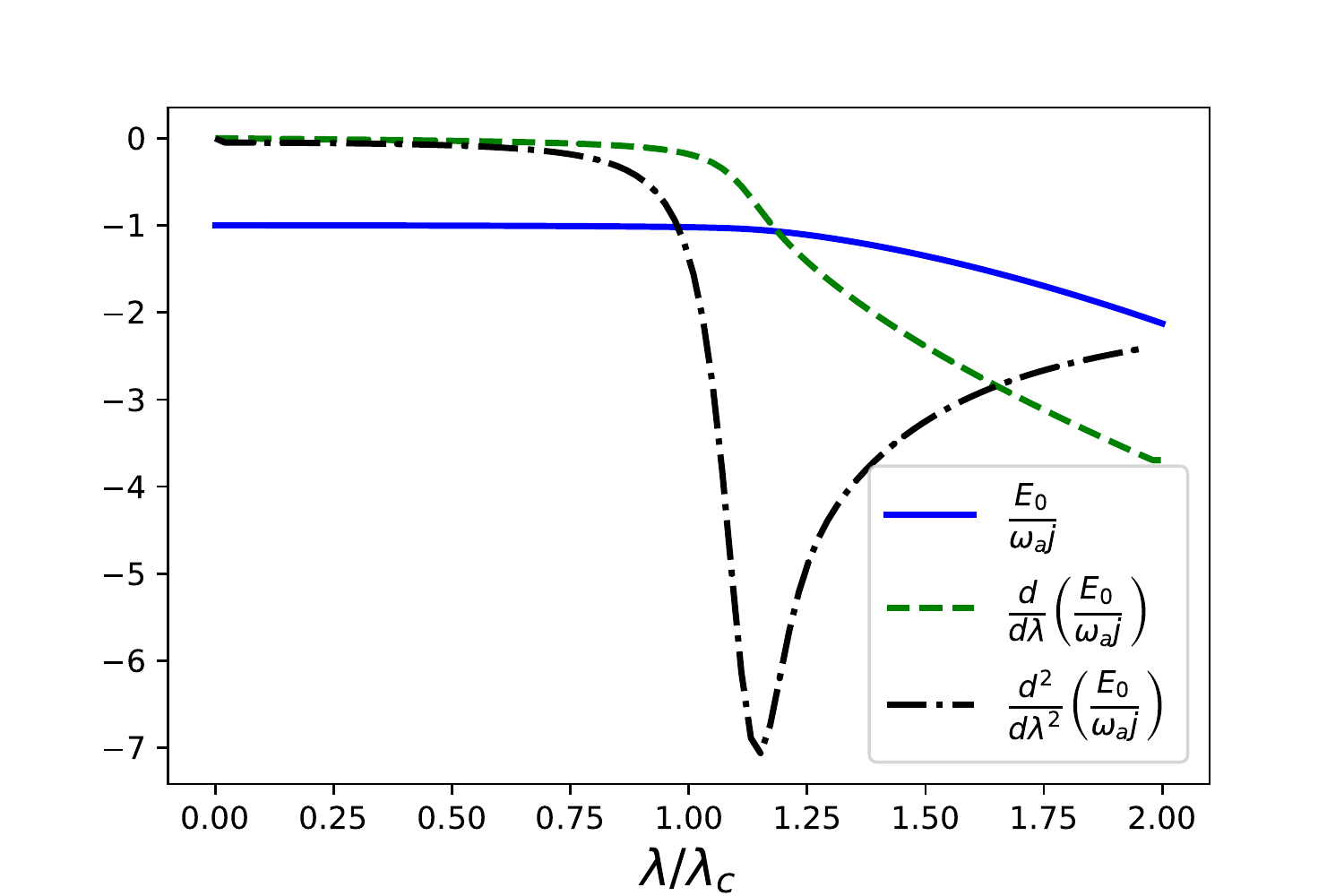}
    \caption{Variation of the scaled ground state, $\frac{E_0}{\omega_a j}$ and its derivatives with the atom-field interaction strength $\lambda$ in case of the $N$-qubit Dicke model (Eq. (\ref{eq-DickeHam})). Here we have chosen $N=20, \text{and }~\omega_a=\omega_c = 2$.}
    \label{fig-Dicke_Ham_Phase_Transition_with_derivatives}
\end{figure}

It consists of $N$ two-level atoms with transition frequencies $\omega_a$, coupled to a single mode of a quantized radiation field of frequency $\omega_c$. This can be derived from the generalized Dicke model in Sec. \ref{sec-gen_Dicke_model}, for equal interaction strengths of both co- and counter-rotating terms. The Hamiltonian (for $\hbar = 1$) of the system is given by
\begin{equation}
    \label{eq-DickeHam}
    H_D = \omega_a J_z + \omega_c \hat{a}^\dagger \hat{a} + \frac{\lambda}{\sqrt{N}} (J_+ + J_-)(\hat{a} + \hat{a}^{\dagger}).
\end{equation} 
The $N$-qubit Dicke model (Eq. (\ref{eq-DickeHam})) undergoes a superradiant phase transition~\cite{Hepp_Leib2, Hioe2} when the atom-field interaction strength reaches a critical value of $\lambda_c = \frac{\sqrt{\omega_a\omega_c}}{2}$. This model is symmetric under the parity operator $\Pi = \exp\left[-i\pi\left(\hat a^\dagger \hat a + J_z\right)\right]$, and supports a $\mathbb{Z}_2$ symmetry. It is this symmetry that breaks in the superradiant phase at the critical value of $\lambda_c$. The phase transition in the ground state ($E_0$) of the Dicke model is shown in Fig. (\ref{fig-Dicke_Ham_Phase_Transition_with_derivatives}). Here, we can observe a discontinuity in the second-order derivative of the ground state of the Hamiltonian, indicating a second-order phase transition around the critical point $\lambda_c$. This critical limit of the phase transition in the $N$-qubit Dicke model can also be derived from Eq. (\ref{eq-critical_phase_trans}). In the limit $\lambda \to \lambda'$, the value of $R$ goes to 0 and the critical coupling reduces to $\lambda_c^2 = -\frac{(\omega_c^2 + \kappa^2)(\omega_a^2 + \gamma^2)}{8\omega_a\omega_c s_z}$. Further, under the condition of zero dissipation ($\gamma = \kappa = 0$) and zero external temperature, for a fully polarized initial state ($s_z = -1/2$), this critical value becomes $\lambda_c = \frac{\sqrt{\omega_a\omega_c}}{2}$.
\subsubsection{Tavis-Cummings (TC) model}\label{sec-TC_model}
\begin{figure}[!htb]
    \centering
    \includegraphics[width = 1\columnwidth]{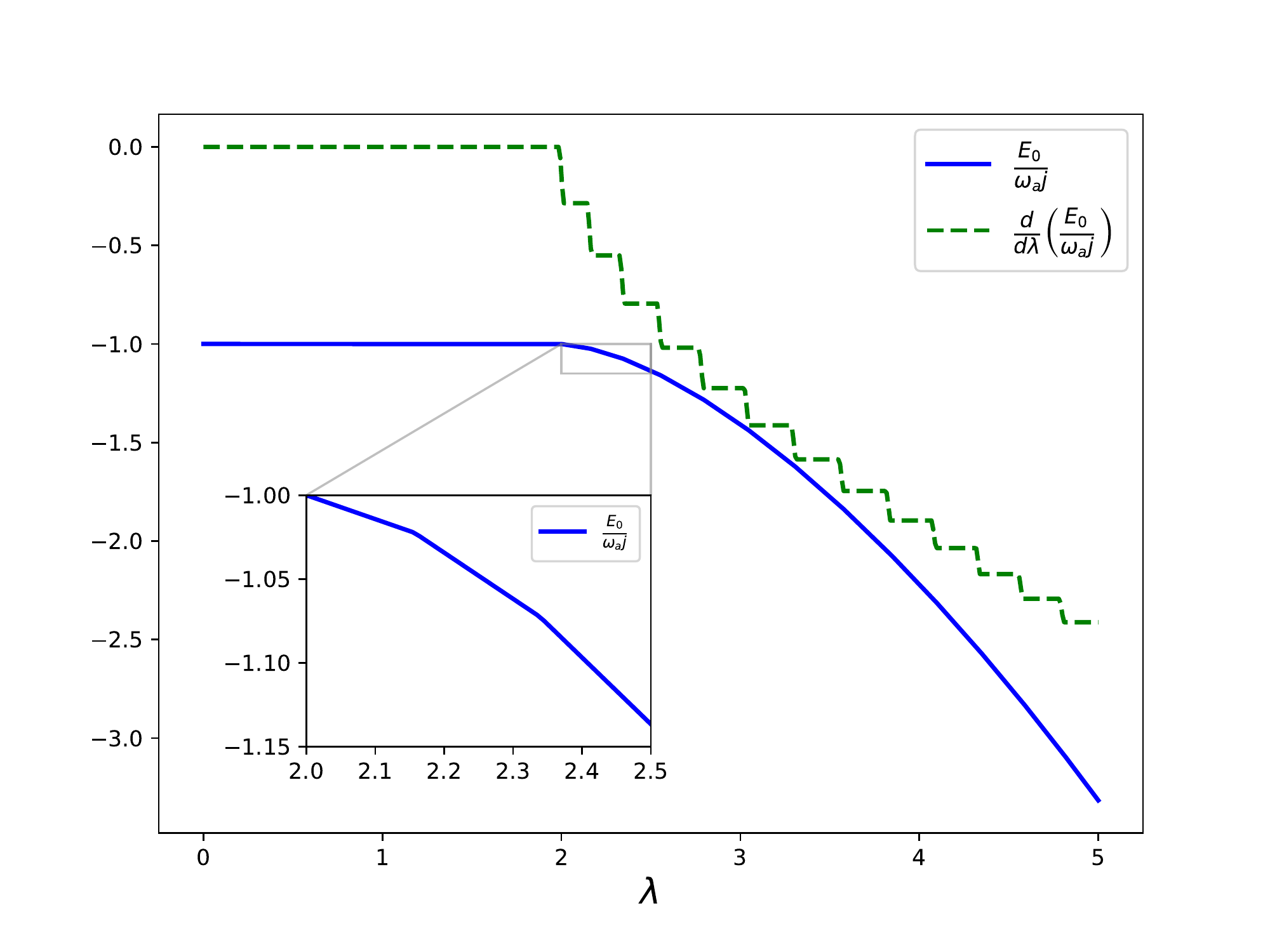}
    \caption{Variation of the scaled ground state energy $\frac{E_0}{\omega_a j}$ and its derivatives with the atom-field interaction strength $\lambda$ for the TC model. The inset plot shows the variation of the ground state energy $\frac{E_0}{\omega_a j}$ zoomed in between $\lambda = 2$ and 2.5.} The parameters, in this case, are $N = 7$, $\omega_a = \omega_c = 2$.
    \label{fig-TC_model_Phase_Transition_with_derivatives}
\end{figure}

The Tavis-Cummings (TC) model can be arrived at after removing the counter-rotating terms from the generalized Dicke model (Eq. (\ref{eq-gen_dicke_Ham})). The Hamiltonian of the TC model is given as
\begin{equation}
    H_{\text{TC}} = \omega_c \hat{a}^\dagger\hat{a} + \frac{\omega_a}{2} \sum_{j = 1}^N \sigma_j^z + \frac{\lambda}{\sqrt{N}}\sum_{j = 1}^N\left(\sigma^+_j\hat{a} + \sigma^-_j\hat{a}^\dagger \right).
\end{equation}
The total number of excitations $\mathcal{N} = \hat{a}^\dagger\hat{a} + \frac{1}{2}\sum_j\sigma^z_j$ in this model remain conserved.
Using collective angular momentum operators, we can rewrite the TC Hamiltonian as 
\begin{equation}
    H_{\text{TC}} = \omega_c \hat{a}^\dagger\hat{a} + \omega_a J_z +  \frac{\lambda}{\sqrt{N}} \left(J_+\hat a + J_- \hat a^\dagger\right).
    \label{eq-TC_modelHam}
\end{equation}
The TC model was shown to undergo a phase transition at a critical value of atom-cavity field interaction strength $\lambda_c = \sqrt{\omega_a\omega_c}$~\cite{HEPP_Lieb, wang-hioe} at zero temperature. This is 2 times the critical coupling obtained for the $N$-qubit Dicke model because of the presence of rotating wave approximation in this model. This phase transition can be observed in the ground state of the TC model as shown in Fig. \ref{fig-TC_model_Phase_Transition_with_derivatives}. Here, we can see that for the given parameters $\omega_a = \omega_c = 2$, the ground state of the TC model becomes non-analytical after $\lambda = 2$. Interestingly, the TC model shows both first- and second-order phase transitions. The case considered here, with seven atoms in the TC model, shows a first-order phase transition. Further, the first-order derivative of the ground state energy has a step-like discontinuity, which is consistent with the observations in~\cite{Larson_2017, Castanos_2009}. This step-like discontinuity arises due to sudden changes in the slope of the ground state energy function, as depicted in the inset plot of Fig. ~\ref{fig-TC_model_Phase_Transition_with_derivatives}. The derivative of the ground state energy becomes a continuous function when we consider the thermodynamic limit ($N \to \infty$) of the number of atoms. In the thermodynamic limit, the second-order derivative turns out to be discontinuous, depicting a second-order phase transition. However, under dissipation, when we consider the losses due to atomic system and cavity decay, the criticality of the TC model vanishes~\cite{Larson_2017}. This can also be seen from Eq. (\ref{eq-critical_phase_trans}), where for $\lambda' = 0$ (i.e., $R = 1$), the critical coupling $\lambda_c$ does not have real solutions.
\subsubsection{Floquet Dicke model}\label{sec-floquet-dicke-model}
\begin{figure}[!htb]
    \centering
    \includegraphics[width = 1\columnwidth]{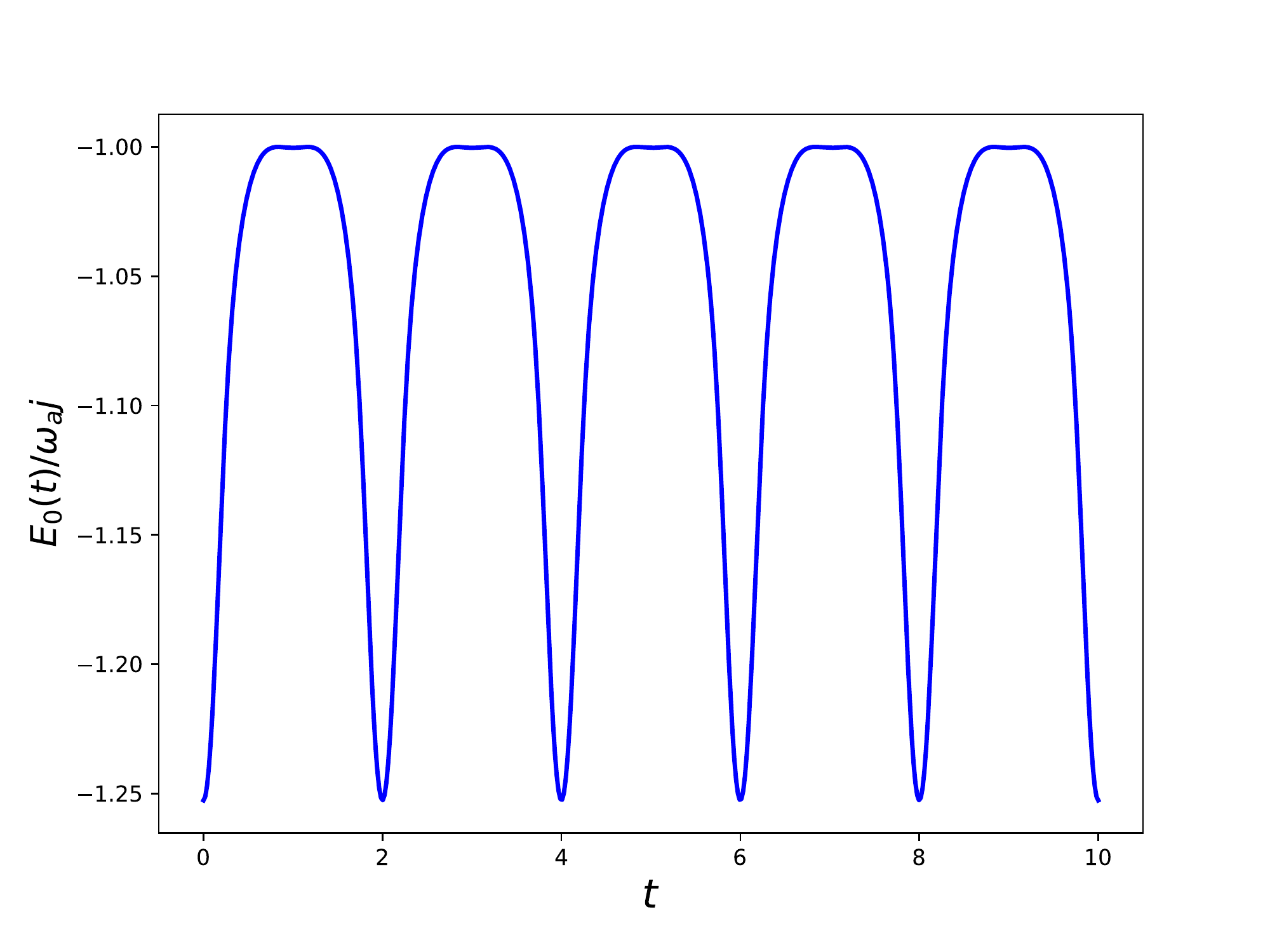}
    \caption{Variation of the scaled ground state, $\frac{E_0}{\omega_a j}$ with time $t$ for the Floquet Dicke model. Here we have chosen $N=7$, $\omega_a = \omega_c = 2$, $\lambda_0 = 0.65$, $\Delta \lambda = 0.75$, and $\Omega = \pi$.}
    \label{fig-Floquet_Dicke_model_ground_state_with_t}
\end{figure}

In the Floquet Dicke (FD) model, we replace the coupling between the atom and the cavity field with a time-dependent one; that is, we replace the value of $\lambda$ in Eq. (\ref{eq-DickeHam}) with, for example,  $\lambda(t) = \lambda_0 + \Delta\lambda\cos(\Omega t)$. Thus, the Hamiltonian for this model can be explicitly given as 
\begin{equation}
   \label{eq-Floquet_DickeHam}
     H_{FD} = \omega_a J_z + \omega_c \hat{a}^\dagger \hat{a} + \frac{\lambda(t)}{\sqrt{N}} (J_+ + J_-)(\hat{a} + \hat{a}^{\dagger}), 
\end{equation}
where $J_{i}\{i = x, y, z\}$ are the collective angular momentum operators, and $\hat a~(\hat a^\dagger)$ are the bosonic annihilation (creation) operators. It is opportune here to discuss the phase transition tendencies of this model. In Fig. \ref{fig-Floquet_Dicke_model_ground_state_with_t}, we can observe the pattern of the scaled ground state energy of the Floquet Dicke model, varying in time. With values of $\lambda_0 = 0.65$ and $\Delta\lambda=0.75$, the superradiant phase transition in the FD model is evident when the time-varying interaction strength crosses the critical value of $\sqrt{\omega_a\omega_c}/2 = 1$. The value of the ground state energy is constant at -1.0 in the normal phase, and it decreases quickly as the phase becomes superradiant. Therefore, the system oscillates between the normal and the superradiant phases. 

The numerical simulation of the various Dicke models is important as they have an infinite bosonic Fock space. This entails a truncation of the bosonic Fock space. The scheme of truncation depends on the observable we want to calculate. For example, if we want to calculate the ground state energy $E_0$ of the $N$-qubit Dicke model, we arbitrarily choose the dimension $n$ of the bosonic Fock space. We then calculate the difference $E_0(n+1) - E_0(n)$. If this difference is significantly small, we fix the value of $n$. However, in the case of the strong coupling regime, the dimension of the Hilbert space increases very rapidly, and for large system sizes the complexity of the calculation increases manifold. To tackle this issue and also to find an exact numerical solution, particularly in the case of the $N$-qubit Dicke model, the formalism of the extended bosonic coherent basis (or efficient basis) was developed in~\cite{Chen_exact1}, and further elucidated in~\cite{magnani_Hirsch_efficient1, magnani_Hirsch_efficient2}. The idea of the conserved parity operator that commutes with the $N$-qubit Dicke model was used to arrive at this basis. Using this, the calculation of the observables converges in the case of the $N$-qubit Dicke model for a higher number of atoms and in a strong coupling regime.
\subsection{Out-of-time-ordered-correlator (OTOC)}\label{sec-OTOC}
The out-of-time-ordered correlator (OTOC) is defined as
\begin{equation}
    \label{eq-otoc1}
    \mathcal{C}_{AB}(t) = \braket{[A_t, B]^\dagger[A_t, B]}, 
\end{equation}
where $A_t$ and $B$ are arbitrary Heisenberg operators in the Hilbert space. $\braket{\cdot} = {\rm Tr}(\rho \cdot)$ denotes the average with respect to the thermal density matrix of the system $\rho = e^{-\beta H}/Z$, where $Z= {\rm Tr}(e^{-\beta H})$ and $\beta = 1/k_B T$. Here $H$ is the Hamiltonian of the system. 
The intuition for $C(t)$ to be connected to chaos comes from the study of the semi-classical limit of a one-particle quantum chaotic system as in~\cite{otoc_Ovchinnikov}. In the semi-classical limit, for $A_t = x(t)$ and $B = p$, we can rewrite the commutator in Eq. (\ref{eq-otoc1}) as the Poisson bracket $i\hbar\{x(t), p\} = i\hbar \frac{\partial x(t)}{\partial x(0)}$. This Poisson bracket is sensitive to the initial conditions and dependent on $e^{\lambda_L t}$ for a classically chaotic system. In the context of quantum chaos, the quasi-classical limit on the growth of OTOC is given by $\mathcal{C}_{AB}(t) \sim \hbar^2e^{2\lambda_L t}$. The exponent $\lambda_L$ is referred to as the Lyapunov exponent. A maximum bound on the Lyapunov exponent $\lambda_L$ was proposed in~\cite{Maldacena2016}, satisfying 
\begin{equation}
    \lambda_L \le \frac{2\pi}{\beta\hbar} = \frac{2\pi k_B T}{\hbar}. 
    \label{eq-lyapunov-inequality}
\end{equation}%
For Hermitian operators $A_t$ and $B$, the form of OTOC defined in Eq. (\ref{eq-otoc1}) becomes, 
\begin{equation}
\label{eq-otoc2}
    \mathcal{C}_{AB}^{\text{herm}}(t) = -\braket{[A_t, B]^2}.
\end{equation}
In general, we have time evolution of the operator $A_t = e^{\frac{i}{\hbar}Ht} A e^{-\frac{i}{\hbar}Ht}$. In the context of open quantum systems, here we make use of the adjoint master equation~\cite{breuer-petruccione, SANDULESCU} for the evolution of the Heisenberg operator $A$ given by 
\begin{align}
    \label{eq-adjoint-mastereq1}
    \frac{dA_t}{dt} &= \mathcal{L}^\dagger A_t,
\end{align}
where $\mathcal{L}^\dagger$ is the superoperator corresponding to the adjoint master equation and can be given by $\mathcal{L}^\dagger (\cdot) = i\left[H, (\cdot)\right] + \sum_k\left[\frac{\nu_k}{2}\left(N_{th,k} + 1\right)\left(2L_k^\dagger (\cdot) L_k - \left\{L_k^\dagger L_k, (\cdot)\right\}\right)\right] + \sum_k\left[\frac{\nu_k N_{th,k}}{2}\left(2L_k (\cdot) L_k^\dagger - \left\{L_k L_k^\dagger, (\cdot)\right\}\right)\right]$, where $H$ is the Hamiltonian of the respective model and $N_{th, k} = \frac{1}{e^{\beta\hbar\omega_k} - 1}$. In the case of the Dicke class of models, we consider the following decay rates: $i).$ decay due to the spontaneous emission of the atom, and $ii).$ leakage of photons from the cavity; this is detailed below in Eq. (\ref{eq-adjoint_master_eq_qt}). $L_k(L_k^\dagger)$ are the Lindblad operators. 
The form of OTOC given in Eq. (\ref{eq-otoc1}), in a generic case, can be expressed as
\begin{equation}
    \label{eq-otoc3}
    \mathcal{C}_{AB}(t) = \mathcal{D}_{AB}(t) + \mathcal{I}_{AB}(t) - 2\Re\left\{\mathcal{F}_{AB}(t)\right\}, 
\end{equation}
where $\mathcal{D}_{AB}(t) = \braket{B^\dagger (A^\dagger A)_t B}$, $\mathcal{I}_{AB} = \braket{A_t^\dagger (B^\dagger B)A_t}$, and $\mathcal{F}_{AB}(t) = \braket{A_t^\dagger B^\dagger A_t B}$. $\mathcal{I_{AB}}$ and $\mathcal{D_{AB}}$ are the time-ordered correlation functions, and $\mathcal{F}_{AB}$ is the genuine out-of-time-ordered four-point correlation function. This is the reason why $\mathcal{F_{AB}}$ is commonly referred to as OTOC. 
The $\mathcal{D}_{AB}(t), \mathcal{C}_{AB}(t)$, and $\mathcal{I_{AB}} (t)$ are connected to each other via the following relation~\cite{ueda_3},
\begin{equation}
    \label{eq-relation_C_I_D}
    \frac{\mathcal{C}_{AB}}{\mathcal{I_{AB}}} = \left(1 + \alpha_t \sqrt{\frac{\mathcal{D}_{AB}}{\mathcal{I_{AB}}}}\right)^2, 
\end{equation}
where $\alpha_t$ is a time-dependent numerical factor satisfying $|\alpha_t|\le 1$. 
It can also be seen that in the case where $A$ and $B$ are unitary operators, Eq. (\ref{eq-otoc3}) can be rewritten as
\begin{equation}
    \label{eq-otoc4}
    \mathcal{C}_{AB}^{\text{uni}}(t) = 2\left(1 - \Re\left\{\mathcal{F}_{AB}(t)\right\}\right). 
\end{equation}
Also, for Hermitian operators $A_t$ and $B$, the function $\mathcal{F}_{AB}(t)$ has a uniform asymptotic expansion given by 
\begin{equation}
    \mathcal{F}_{AB}(t) = c_0 - \epsilon c_1 e^{\lambda_L t} + O(\epsilon^2). 
    \label{eq-f_ab_asymptotic_expansion}
\end{equation}

In many cases, instead of Eq. (\ref{eq-otoc1}), a regularized form is adopted for the definition of OTOC. This is achieved by splitting the thermal density matrices $\rho$ into two $\rho^{1/2}$'s, one inserted between the commutators and the other placed in front of them. The definition is given as 
\begin{equation}
    \label{eq-otoc5}
    \mathcal{C}_{AB}^{\text{reg}}(t) = {\rm Tr} \left\{\rho^{1/2}[A_t, B]^\dagger \rho^{1/2} [A_t, B]\right\}.
\end{equation}
In~\cite{ueda1}, the OTOC was defined in terms of both commutator and anti-commutator for Hermitian operators $A$ and $B$. The regularized OTOC is redefined in terms of both (anti-)commutators as 
\begin{equation}
    \label{eq-regOTOC2}
    \mathcal{C}_{AB}^{\text{reg}}(x, y, t, t') = \mathcal{C}_{[A, B]_x[A, B]_y}^{\text{reg}} (t, t') = {\rm Tr}\left(\rho^{1/2}[A_t, B_{t'}]_x\rho^{1/2}[A_t, B_{t'}]_y\right),
\end{equation}
where $x, y$ can have values $\pm$, and $[.]_{(-)+}$ represents the (anti-)commutator. The negative of the ordinary OTOC (Eq. (\ref{eq-otoc2})) for Hermitian operators $A$ and $B$ in the form of expectation value ${\rm Tr}(\rho \cdots)$ of products of (anti-)commutator for a given state $\rho$ is given as
\begin{equation}
    \label{eq-physOTOC}
    \mathcal{C}_{AB}^{\text{phys}} (x, y, t, t') = \mathcal{C}_{[A, B]_x[A, B]_y}^{\text{phys}} (t, t') = {\rm Tr}\left(\rho[A_t, B_{t'}]_x[A_t, B_{t'}]_y\right).
\end{equation}
We shall refer to the form of OTOC given above as a physical OTOC, and $\mathcal{C}_{[A, B]_x[A, B]_y}^{\text{phys}} (t) = -\mathcal{C}_{AB}^{\text{herm}} (t)$ for x= y = +. The difference between the regularized OTOC and the physical OTOC is given by the Wigner Yanase (WY) Skew information, 
\begin{equation}
    \label{eq-WYSkewInf}
    I_{1/2}(\rho, O) = -\frac{1}{2}{\rm Tr}\left([\rho^{1/2}, O]^2\right) = {\rm Tr}\left(\rho O^2\right) - {\rm Tr}\left(\rho^{1/2}O\rho^{1/2}O\right),
\end{equation}
for a Hermitian operator $O$. Using the WY skew information, the relation between the regularized and physical OTOCs is written as
\begin{align}
    \label{eq-Relation-reg_physOTOC1}
    \mathcal{C}_{[A, B]^2}^{\text{reg}} (t, t') &= \mathcal{C}_{[A, B]^2}^{\text{phys}} (t, t') + I_{1/2} (\rho, i[A_t, B_{t'}]), \\
    \label{eq-Relation-reg_physOTOC2}
    \mathcal{C}_{\{A, B\}^2}^{\text{reg}} (t, t') &= \mathcal{C}_{\{A, B\}^2}^{\text{phys}} (t, t') - I_{1/2} (\rho, \{A_t, B_{t'}\}), \\
    \label{eq-Relation-reg_physOTOC3}
    \mathcal{C}_{\{A, B\}[A, B]}^{\text{reg}} (t, t') &= \frac{1}{2} \left[\mathcal{C}_{\{A, B\}[A, B]}^{\text{phys}} (t, t') + \mathcal{C}_{[A, B]\{A, B\}}^{\text{phys}} (t, t')\right] \nonumber \\
    &+ \frac{i}{4} I_{1/2}\left(\rho, \{A_t, B_{t'}\} + i[A_t, B_{t'}]\right) \nonumber \\ &-\frac{i}{4} I_{1/2}\left(\rho, \{A_t, B_{t'}\} - i[A_t, B_{t'}]\right). 
\end{align}
The physical and regularized forms of OTOC discussed above establish a relation between the WY Skew information and OTOC. The WY Skew information metric serves as a measure of information contained in quantum fluctuations of observables in the context of quantum information theory and provides a balance between the physical and regularized forms of OTOC. Further, the regularized form of OTOC has been used in a number of works, such as~\cite{Maldacena2016}, and the physical form is defined as the standard expectation value of a quantity. Moreover, the commutator and anti-commutator forms of OTOC are related via an out-of-time-ordered Fluctuation-Dissipation relation~\cite{ueda1} that relates the chaotic behavior of the system to the non-linear response function.
\section{Behavior of Out-of-time-ordered-correlator in quantum optical models}\label{sec-otoc_behavior}
Here, we discuss the evolution of out-of-time-ordered correlator (OTOC) for the quantum optical models discussed in Sec. \ref{sec-quantum_opt_models}, which include the generalized Dicke, $N$-qubit Dicke, Tavis-Cummings, and Floquet Dicke models. Further, we identify, using OTOC, the regime of these quantum optical models showing chaotic behavior. 
\subsection{OTOC in the generalized Dicke model}
\begin{figure}[!htb]
    \centering
    \includegraphics[width = 1\columnwidth]{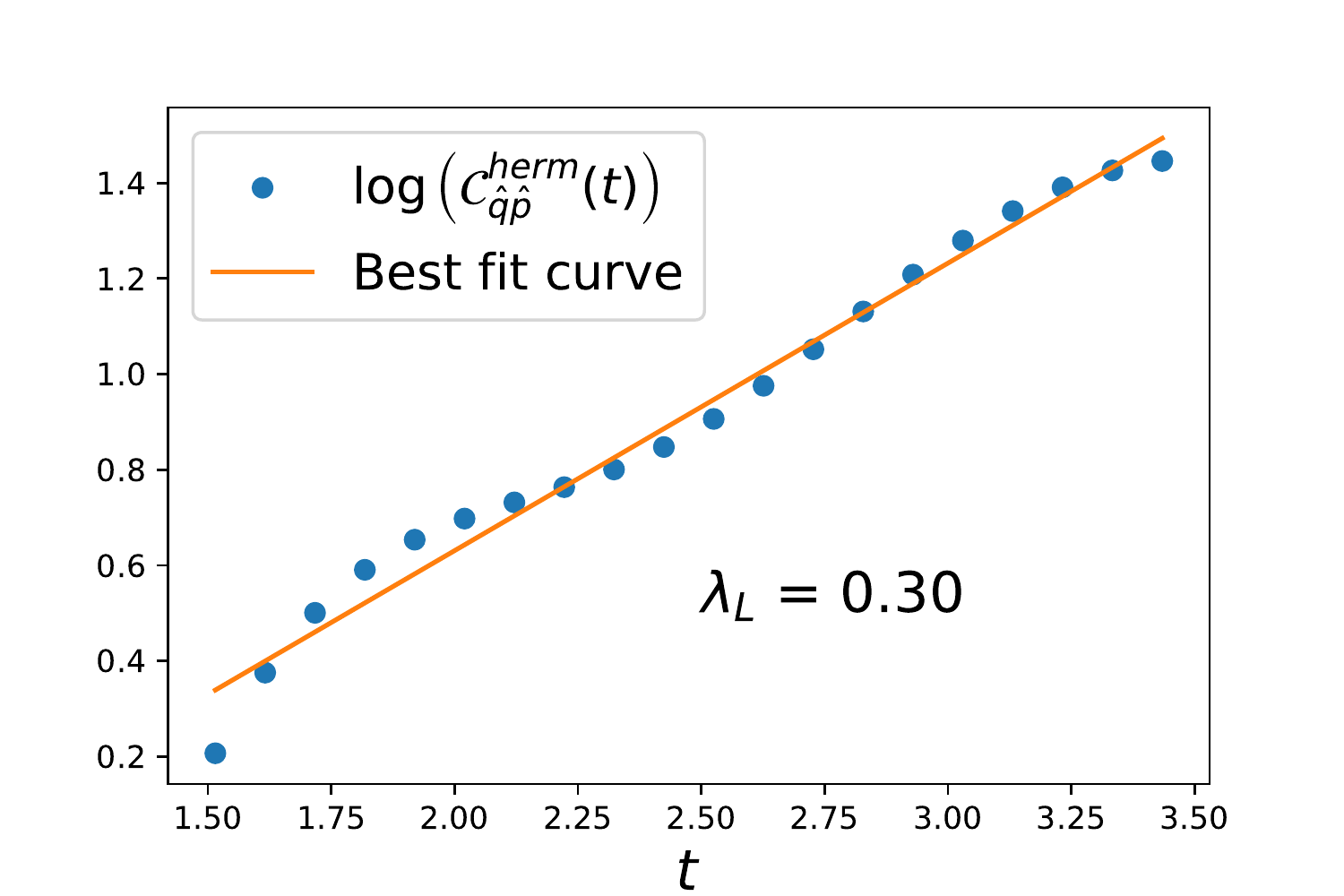}
    \caption{Variation of the OTOC $\log \left(\mathcal{C}_{\hat q\hat p}^{\text{herm}}\right)(t)$ (Eq. (\ref{eq-otoc2})) and its best fit linear curve from $t = 1.5$ to 3.5 for the generalized Dicke model. Here we have chosen $N=7$, $\omega_a = \omega_c = 2$, and $\lambda = 1.5$, $\lambda' = 1.8$, $\gamma = \kappa = 10^{-2}$, and $T = 1$.}
    \label{fig-BestFitCurve_Commutator_gen_Dicke}
\end{figure}

\begin{figure}[!htb]
    \centering
    \includegraphics[width = 1\columnwidth]{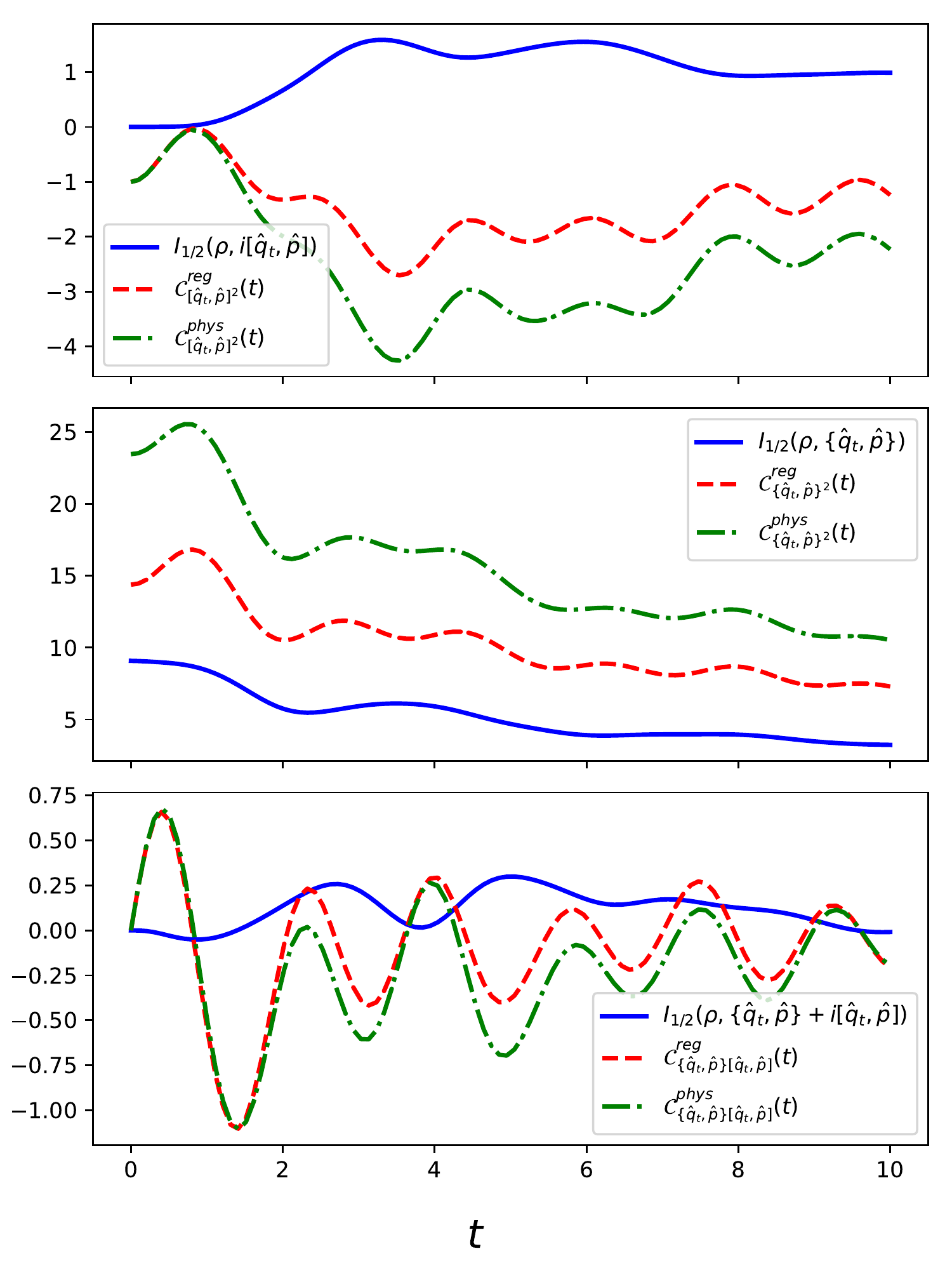}
    \caption{Variation of the various forms of the OTOCs given in Eqs.  (\ref{eq-Relation-reg_physOTOC1}), (\ref{eq-Relation-reg_physOTOC2}), and (\ref{eq-Relation-reg_physOTOC3}), with time for the generalized Dicke model. Here we have chosen $N=7$, $\omega_a = \omega_c = 2$, and $\lambda = 1.5$, $\lambda' = 1.8$, $\gamma = \kappa = 10^{-2}$, and $T = 1$.}
    \label{fig-all_commutator_anti_commutator_otoc_gen_Dicke}
\end{figure}

Here, we study the behavior of the OTOC in the generalized Dicke (GD) model. To this end, we use the hermitian operators $\hat q = (\hat a^\dagger + \hat a)/\sqrt{2}$ and $\hat p = i(\hat a^\dagger - \hat a)/\sqrt{2}$ ($\hat a^\dagger$ and $\hat a$ are the bosonic creation and annihilation operators) in Eq. (\ref{eq-otoc2}), where the operator $\hat q$'s evolution in time is governed by the master equation
\begin{equation}
    \label{eq-adjoint_master_eq_qt}
    \frac{dq_t}{dt} = \mathcal{L}^\dagger(q_t), 
\end{equation}
where the form of $\mathcal{L}^\dagger$ is given in Eq. (\ref{eq-adjoint-mastereq1}). Here, we choose the collective angular momentum operators $J_\pm = J_x \pm iJ_y$ (with $J_{x(y)} = \frac{1}{2}\sum_i\sigma_i^{x(y)}$; $\sigma^{x(y)}$ are the Pauli spin matrices), and the creation and annihilation operators, $\hat a^\dagger,~\hat a$, respectively, as the jump operators. Thus, we have $L_1 = J_-$, and $L_2 = \hat a$ in Eq. (\ref{eq-adjoint-mastereq1}), and $\nu_1 = \gamma$ and $\nu_2 = \kappa$ are the decay rates due to spontaneous emission and leakage of photons from the cavity, respectively. 
We will now make use of the solution of this equation $q_t$ in the subsequent sections to study the OTOC and its variants. 
\begin{figure}[!htb]
    \centering
    \includegraphics[width = 1\columnwidth]{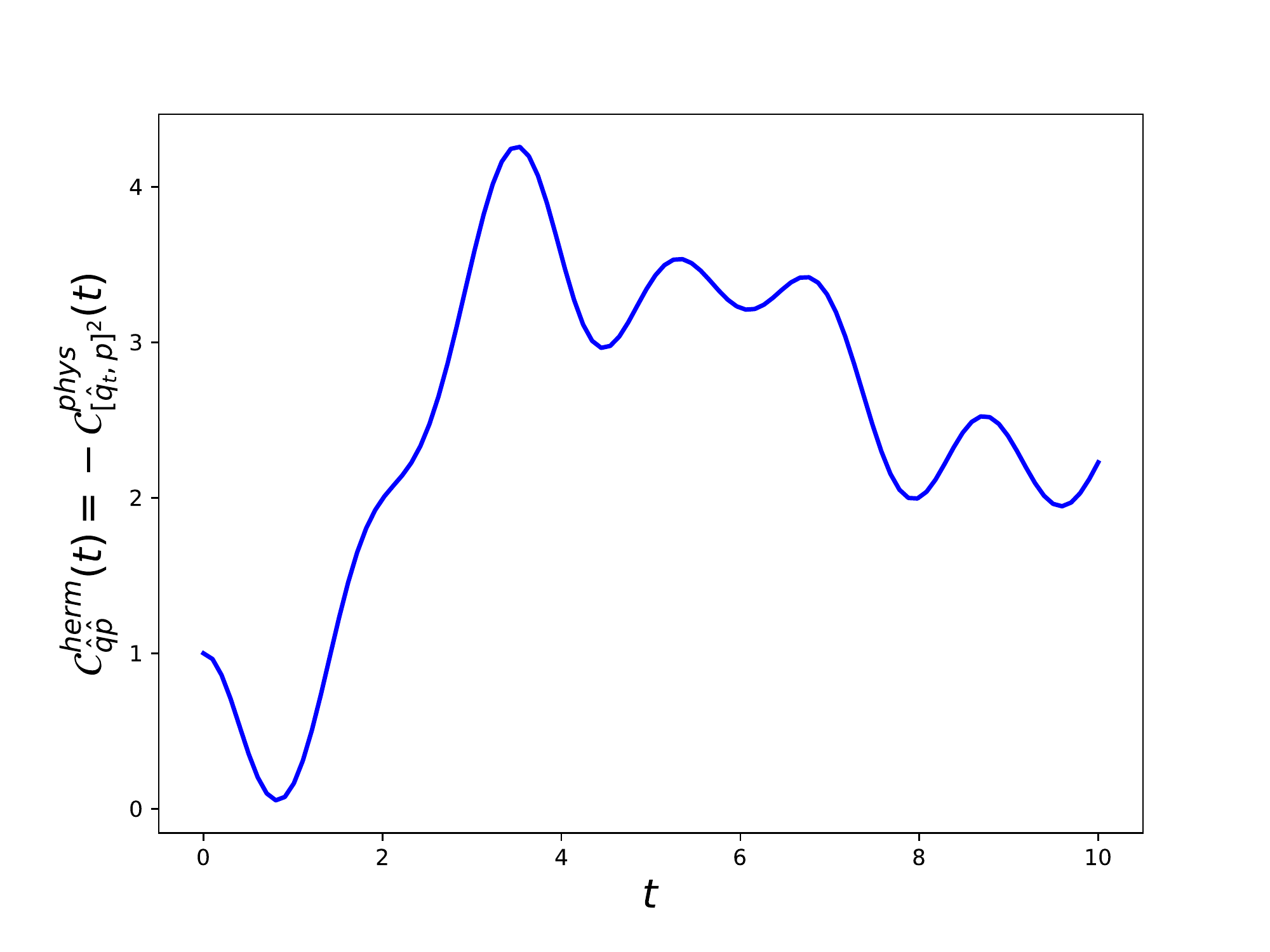}
    \caption{Variation of the OTOC $\mathcal{C}_{\hat q\hat p}^{\text{herm}}(t)$ (Eq. (\ref{eq-otoc2})) or the negative of physical OTOC in commutator form $-\mathcal{C}^{\text{phys}}_{[\hat q_t, \hat p]^2} (t)$, with time for the generalized Dicke model. Here we have chosen $N=7$, $\omega_a = \omega_c = 2$, $\lambda = 1.5$, $\lambda' = 1.8$, $\gamma = \kappa = 10^{-2}$, and $T = 1$.}
    \label{fig-physical_hermitian_otoc_gen_Dicke}
\end{figure}
In Fig. \ref{fig-physical_hermitian_otoc_gen_Dicke}, we can observe the variation of the physical OTOC with time. We have chosen both $\lambda$ and $\lambda'$ here to be in the superradiant phase. An exponential rise in the OTOC can be observed from time $t = 1.5$ to 3.5.
We can plot this region on a logarithmic scale to get the Lyapunov exponent, see above Eq. (\ref{eq-lyapunov-inequality}). This is depicted in Fig. \ref{fig-BestFitCurve_Commutator_gen_Dicke}.
The Lyapunov exponent comes out to be 0.30 in the region from $t=1.5$ to 3.5. This value of the Lyapunov exponent also satisfies the inequality given in Eq. (\ref{eq-lyapunov-inequality}) for $T = 1$ in the natural units where $\hbar = k_B = 1$. We now further observe the variation of the other forms of OTOC given in Eqs. (\ref{eq-regOTOC2}) and (\ref{eq-physOTOC}), together with the Wigner-Yanase skew information function given in Eq. (\ref{eq-WYSkewInf}) to verify the Eqs. (\ref{eq-Relation-reg_physOTOC1}), (\ref{eq-Relation-reg_physOTOC2}), and (\ref{eq-Relation-reg_physOTOC3}). This is depicted in Fig. \ref{fig-all_commutator_anti_commutator_otoc_gen_Dicke}.
In this figure, we can observe that WY Skew information acts as a balance between the regularized and the physical OTOCs for the commutator, anti-commutator, and a mix of (anti-)commutator forms. 

\begin{figure}[!htb]
    \centering
    \includegraphics[width = 1\columnwidth]{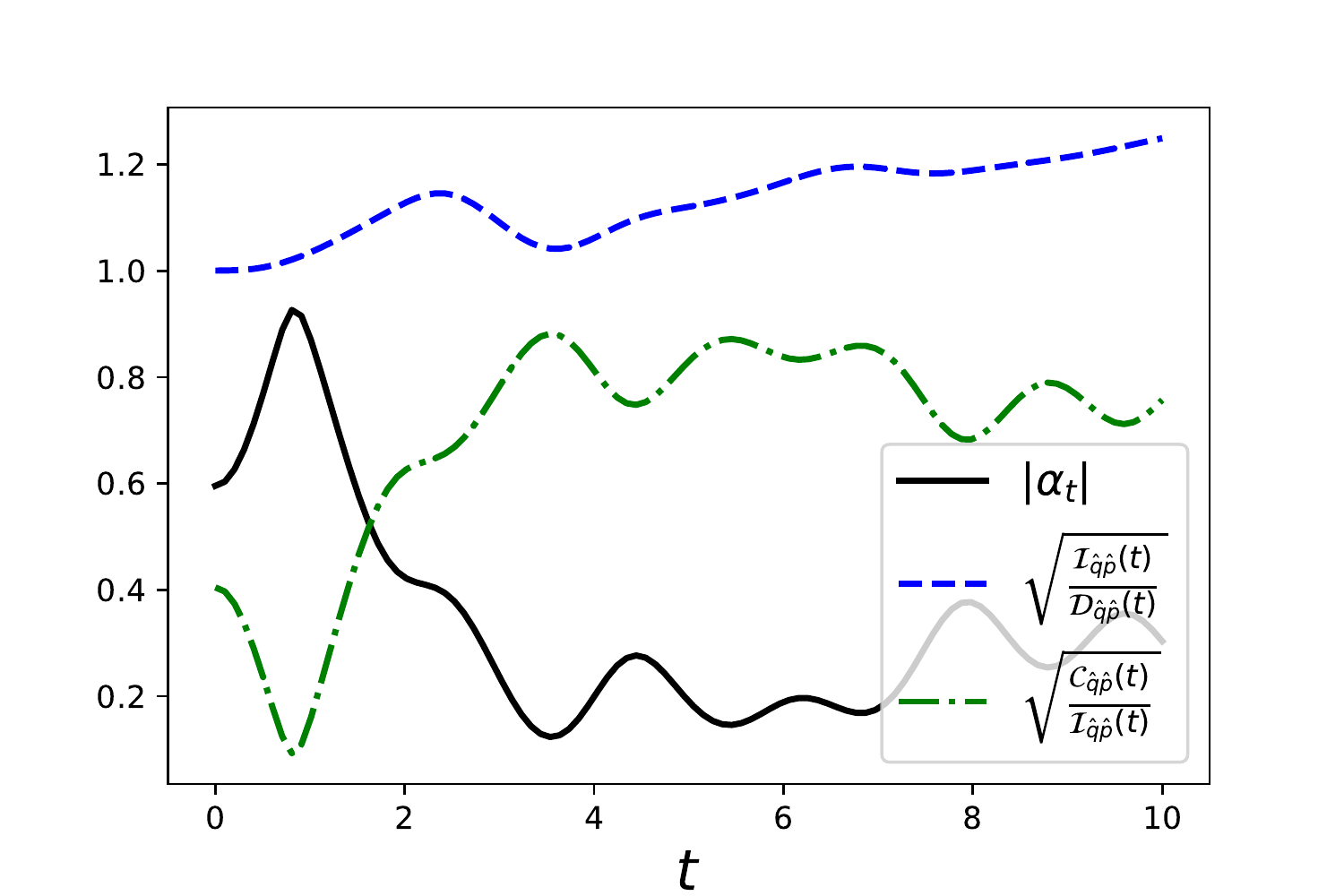}
    \caption{Variation of the parameter $|\alpha_t|$, and the functions $\sqrt{\frac{\mathcal{I}_{\hat q \hat p} (t)}{\mathcal{D}_{\hat q \hat p} (t)}}$ and $\sqrt{\frac{\mathcal{C}_{\hat q \hat p} (t)}{\mathcal{I}_{\hat q \hat p} (t)}}$ given in Eq.~(\ref{eq-alpha_t}), with time in case of generalized Dicke model. Here we have chosen $N=7$, $\omega_a = \omega_c = 2$, and $\lambda = 1.5$, $\lambda' = 1.8$, $\gamma = \kappa = 10^{-2}$, and $T = 1$.}
    \label{fig-variation_alpha_t}
\end{figure}

We further analyze the relation between out-of-time-ordered $\mathcal{C}_{\hat q \hat p}(t)$ and time-ordered functions $\mathcal{D}_{\hat q \hat p}(t)$, and $\mathcal{I}_{\hat q\hat p}(t)$ given in Eq. ({\ref{eq-relation_C_I_D}}). To this end, we can rewrite the time-varying parameter $\alpha_t$ in terms of $\mathcal{C}_{\hat q \hat p}(t)$, $\mathcal{D}_{\hat q \hat p}(t)$, and $\mathcal{I}_{\hat q\hat p}(t)$ as 
\begin{equation}
    \label{eq-alpha_t}
    \alpha_t = \sqrt{\frac{\mathcal{I}_{\hat q\hat p}(t)}{\mathcal{D}_{\hat q\hat p}(t)}}\left(\sqrt{\frac{\mathcal{C}_{\hat q\hat p}(t)}{\mathcal{I}_{\hat q\hat p}(t)}} - 1\right).
\end{equation}
The above relation holds for the value of $\vert \alpha_t \vert\le 1$. The variation of the parameter $\vert \alpha_t \vert$  with time is plotted in Fig.~\ref{fig-variation_alpha_t}.
It is observed that the value of $|\alpha_t|$ lies between zero and one. The pattern of the variation $|\alpha_t|$ is in contrast to the pattern observed for the OTOC in Fig.~\ref{fig-physical_hermitian_otoc_gen_Dicke}. Thus, this behavior of $|\alpha_t|$ decreasing exponentially can point out an exponential increase in the value of OTOC. Further, we can observe the variation of the various ratios of the out-of-time-ordered and time-ordered functions with time in Fig. \ref{fig-variation_alpha_t}. The function $\mathcal{I}_{\hat q \hat p} (t)$ is always greater than the function $\mathcal{D}_{\hat q \hat p} (t)$. Also, initially the function $\mathcal{C}_{\hat q \hat p} (t)$ decreases in comparison to the function $\mathcal{I}_{\hat q \hat p} (t)$. Later it increases to approximately match the value of $\mathcal{I}_{\hat q \hat p} (t)$ as the graph of $\sqrt{\frac{\mathcal{C}_{\hat q \hat p} (t)}{\mathcal{I}_{\hat q \hat p} (t)}}$ can be seen to approach one. 
Moreover, the peaks and valleys of the function $|\alpha_t|$ match exactly with the valleys and peaks of the function $\sqrt{\frac{\mathcal{C}_{\hat q \hat p} (t)}{\mathcal{I}_{\hat q \hat p} (t)}}$. 

\subsection{OTOC in the $N$-qubit Dicke model}\label{sec-n_qubit_Dicke_otoc}
\begin{figure}[!htb]
    \centering
    \includegraphics[width = 1\columnwidth]{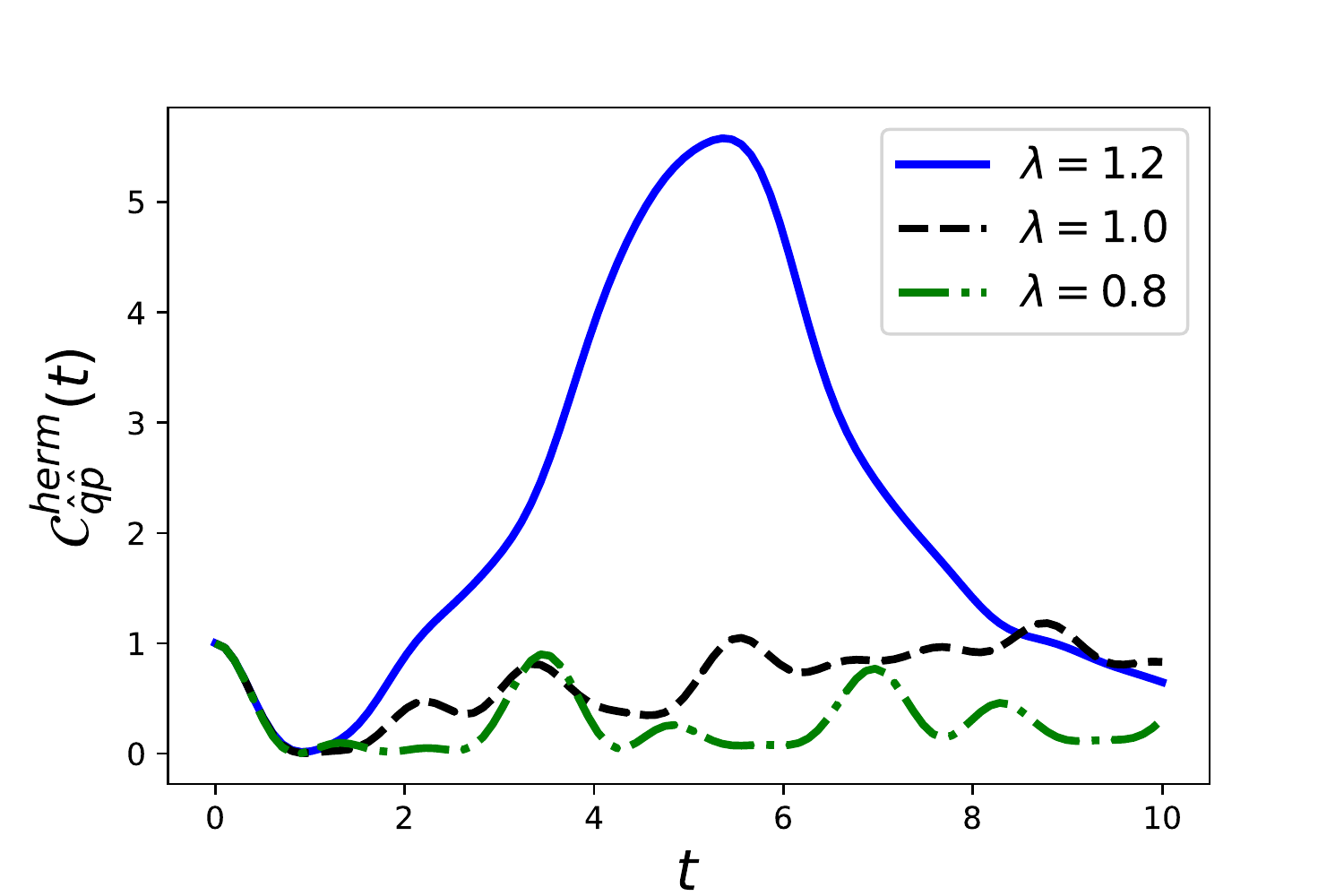}
    \caption{Variation of the OTOC $\mathcal{C}^{\text{herm}}_{\hat q \hat p} (t)$ with time for different values of the spin-cavity system interaction strength $\lambda$ for the $N$-qubit Dicke model. Here we have chosen $N=25$, $\omega_a = \omega_c = 2$, $\gamma = \kappa = 10^{-2}$, and $T = 1$.}
    \label{fig-hermitian_otoc_Dicke_model}
\end{figure}

\begin{figure}[!htb]
    \centering
    \includegraphics[width = 1\columnwidth]{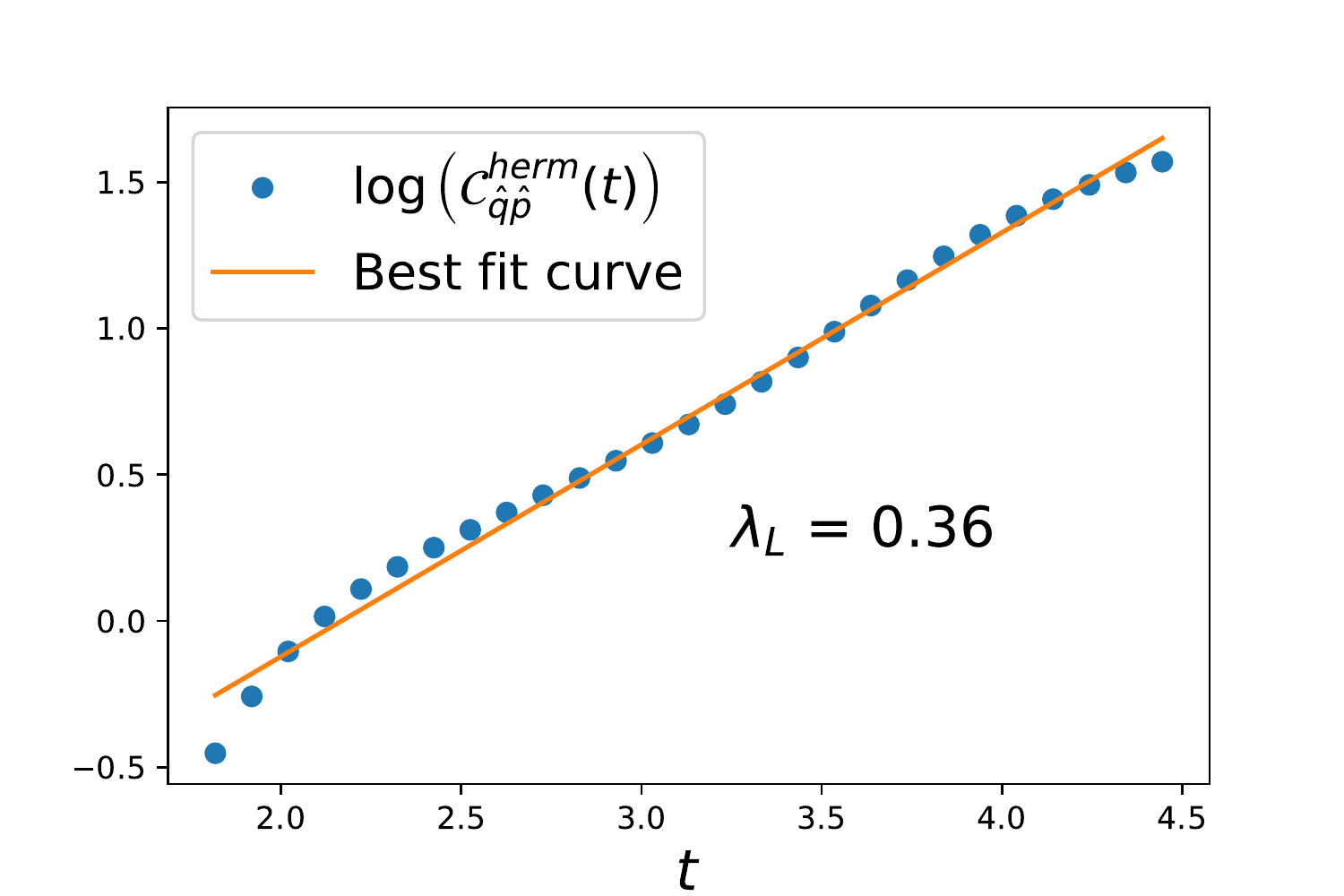}
    \caption{Variation of the OTOC $\log \left(\mathcal{C}_{\hat q\hat p}^{\text{herm}}\right)(t)$ (Eq. (\ref{eq-otoc2})) and its best fit linear curve from $t = 2.0$ to 4.5 for the $N$-qubit Dicke model. Here we have chosen $N=25$, $\omega_a = \omega_c = 2$, $\lambda = 1.2$, $\gamma = \kappa = 10^{-2}$, and $T = 1$.}
    \label{fig-BestFitCurve_Commutator_Dicke_model}
\end{figure}

We now calculate the OTOC for the $N$-qubit Dicke model, introduced in Sec. \ref{sec-N_qubit_Dicke_model}, using the efficient coherent basis introduced in~\cite{Chen_exact1, magnani_Hirsch_efficient1, magnani_Hirsch_efficient2}. In Fig. \ref{fig-hermitian_otoc_Dicke_model}, we can observe the variation of OTOC $\mathcal{C}_{\hat q \hat p} (t)$ for different values of the interaction strength between the spin and the cavity systems $\lambda$. 
It is evident here that the exponential behavior of the OTOC kicks in when the Dicke Hamiltonian is in the superradiant region, i.e., when $\lambda$ is greater than the critical value of phase transition in the Dicke model $\lambda > \lambda_c$ ($\approx1$ for $\gamma = \kappa = 10^{-2}$).  
The Lyapunov exponent is now obtained from the variation of OTOC for the value of $\lambda = 1.2$ using the plot of the function $\log \left(\mathcal{C}_{\hat q\hat p}^{\text{herm}}\right)(t)$ and fitting it with a straight line. This can be seen in Fig. \ref{fig-BestFitCurve_Commutator_Dicke_model} from time $t = 2.0$ to 4.5. In this case, the value of the Lyapunov exponent comes out to be 0.36, which satisfies the inequality given in Eq. (\ref{eq-lyapunov-inequality}) for $T = 1$. 

\begin{figure}[!htb]
    \centering
    \includegraphics[width = 1\columnwidth]{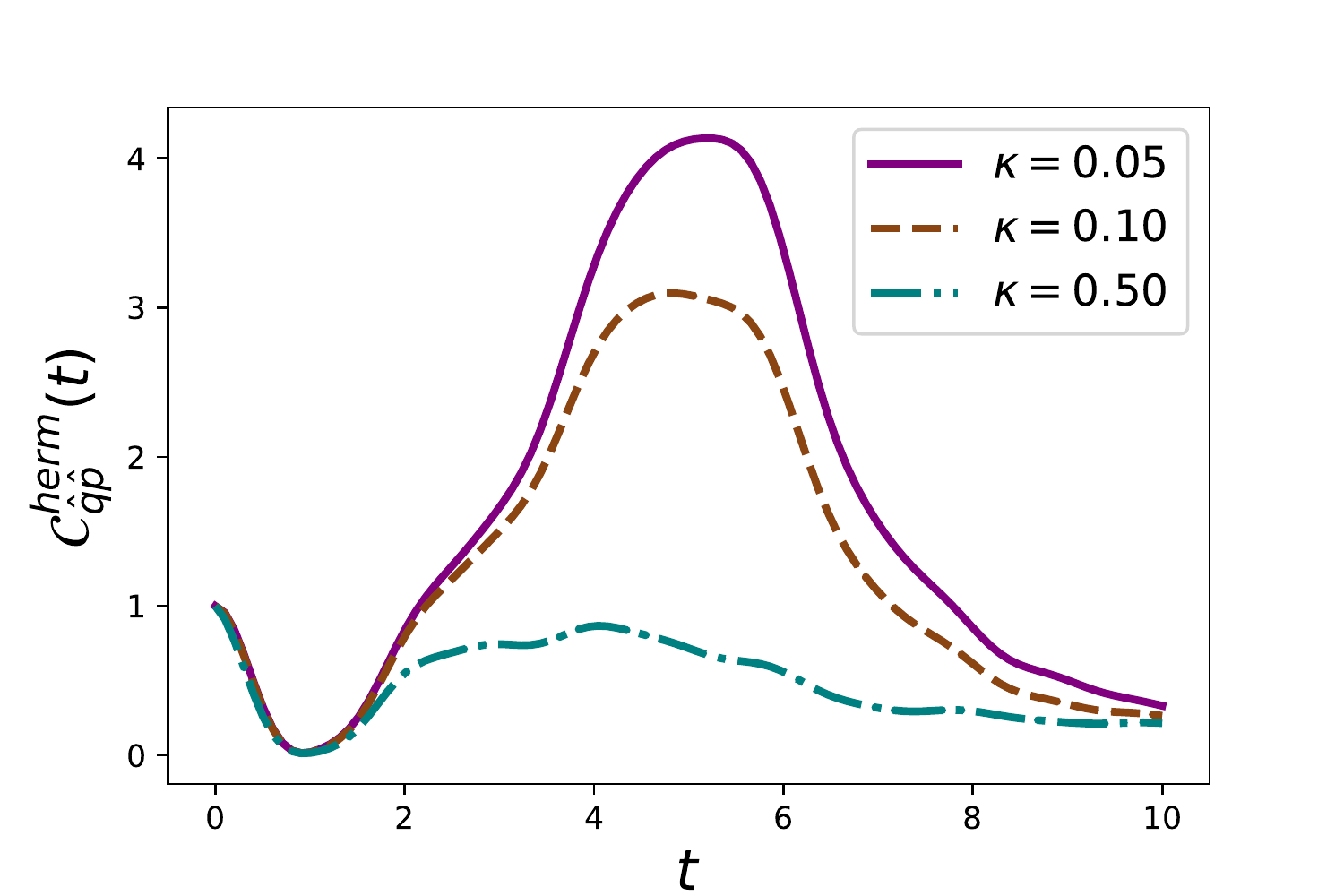}
    \caption{Variation of the OTOC $\mathcal{C}^{\text{herm}}_{\hat q \hat p} (t)$ with time for different values of the dissipative factor $\kappa$ (Eq. (\ref{eq-adjoint_master_eq_qt})), for the $N$-qubit Dicke model. Here we have chosen $N=25$, $\omega_a = \omega_c = 2$, $\lambda = 1.2$, $\gamma = 10^{-2}$, and $T = 1$.}
    \label{fig-hermitian_otoc_Dicke_model_with_gamma}
\end{figure}

Further, we bring out the effect of dissipation on the growth of the OTOC in this model. To this end, for different values of the cavity dissipative factor $\kappa$, see Eq. (\ref{eq-adjoint_master_eq_qt}), we plot the variation of OTOC with time for a given value of spin-cavity system interaction strength $\lambda = 1.2$. This is depicted in Fig. \ref{fig-hermitian_otoc_Dicke_model_with_gamma}.
It is observed that the OTOC is impacted by dissipation. An increment in the value of the cavity dissipative factor $\kappa$ significantly reduces the exponential behavior of the OTOC. For $\kappa = 0.5$, the exponential behavior is completely diminished, and the value of OTOC tends to zero at longer times due to higher dissipation. 

\subsection{OTOC in the Tavis-Cummings model}
\begin{figure}[!htb]
    \centering
    \includegraphics[width = 1\columnwidth]{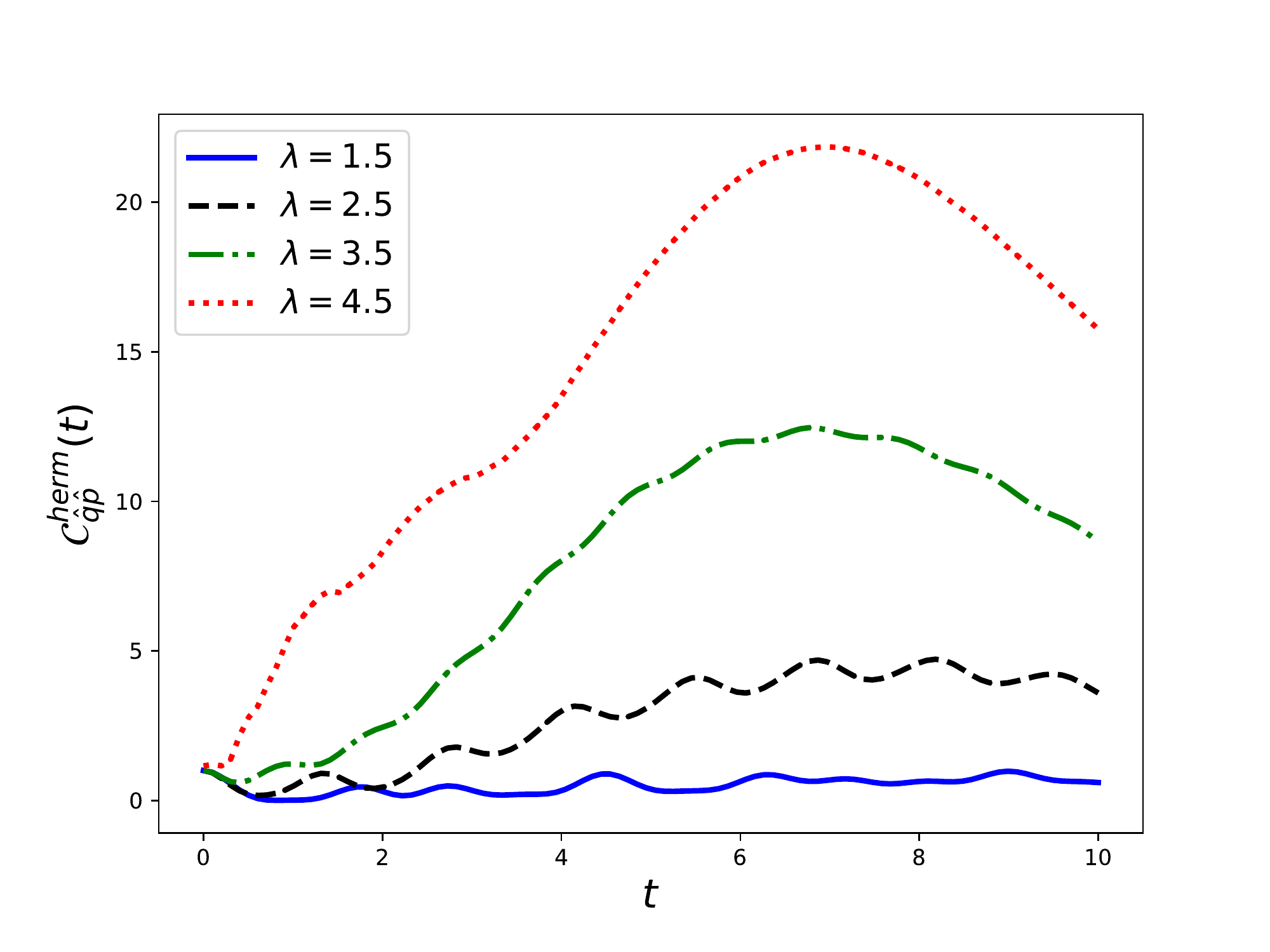}
    \caption{Variation of the OTOC $\mathcal{C}^{\text{herm}}_{\hat q \hat p} (t)$ with time for different values of the spin-cavity interaction strength $\lambda$ in the case of the TC model. Here we have chosen $N=7$, $\omega_a = \omega_c = 2$, $\gamma = \kappa = 10^{-2}$, and $T = 1$.}
    \label{fig-physical_hermitian_OTOC_TC_model}
\end{figure}

Here, we study the behavior of the out-of-time-ordered correlator in the Tavis-Cummings model, discussed in Sec. \ref{sec-TC_model}. The time evolution of the Hermitian OTOC $\mathcal{C}_{\hat q \hat p}^{\text{herm}} (t)$ in this model is depicted in Fig. \ref{fig-physical_hermitian_OTOC_TC_model}. 
The critical value of the system-cavity interaction strength for the phase transition of the TC model for the given parameters is $\lambda_c = 2$ for negligible values of $\gamma$ and $\kappa$, as shown in Fig. \ref{fig-TC_model_Phase_Transition_with_derivatives}. Interestingly, we observe that in this model, the OTOC grows exponentially when
the spin-cavity interaction strength $\lambda$ is greater than the critical value of $\lambda_c = 2$. This brings out the connection of the OTOC with the phase transition, in general, as we observed a similar pattern of OTOC in the case of the generalized Dicke and $N$-qubit Dicke models, where OTOC showed exponential behavior in the superradiant phase characterized by second-order phase transition. 

The Lyapunov exponent can be found using curve-fitting, as done previously for the generalized and the $N$-qubit Dicke models. It comes out to be 0.12 for variation of OTOC $\mathcal{C}^{\text{herm}}_{\hat q \hat p} (t)$ at the value of interaction strength $\lambda = 4.5$ in Fig. \ref{fig-physical_hermitian_OTOC_TC_model}.  
  
\subsection{OTOC in the Floquet Dicke model}
\begin{figure}[!htb]
    \centering
    \includegraphics[width = 1\columnwidth]{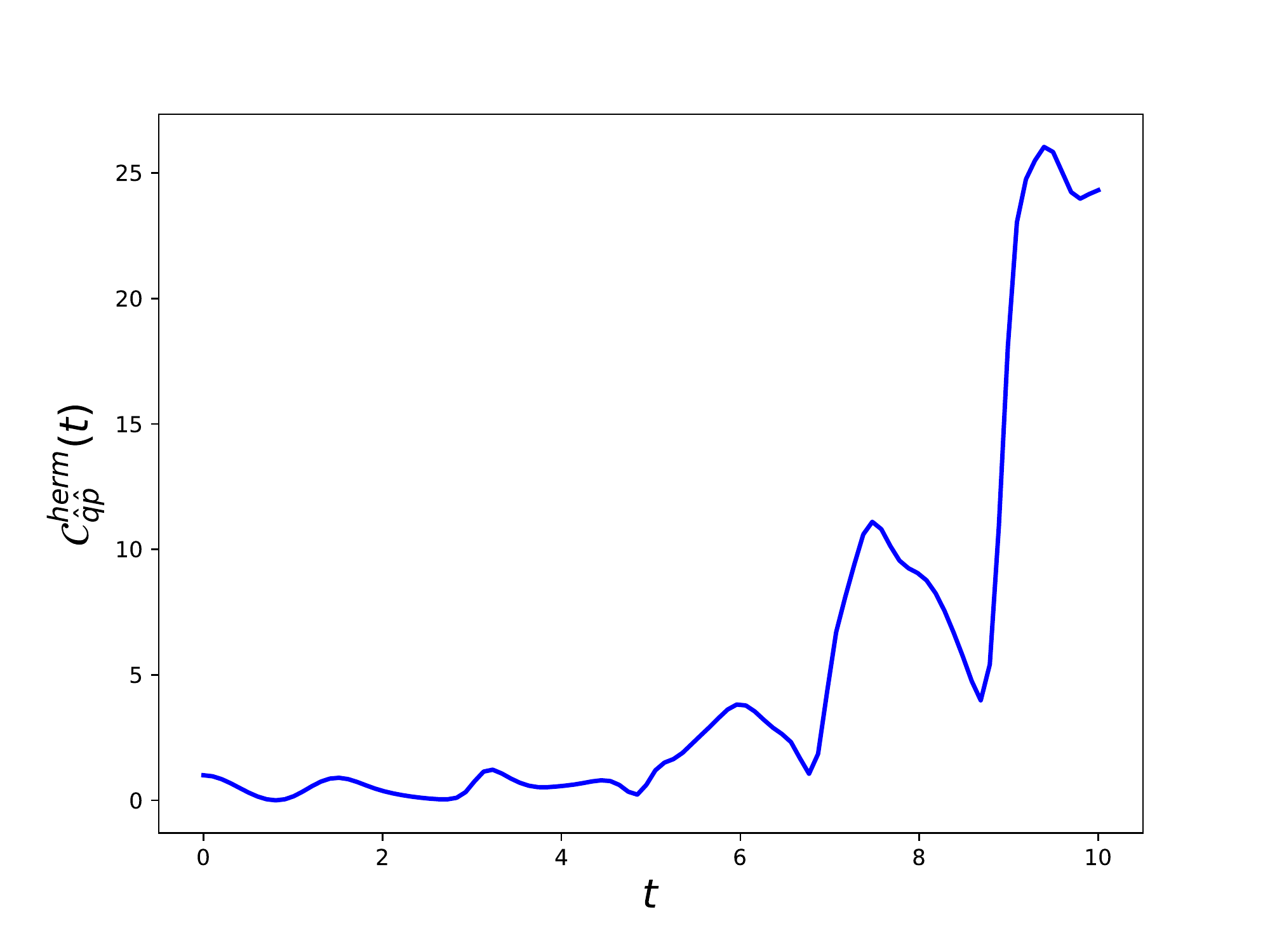}
    \caption{Variation of the OTOC $\mathcal{C}^{\text{herm}}_{\hat q \hat p} (t)$ with time for the Floquet Dicke model. Here we have chosen $N=7$, $\lambda = 0.65$, $\Delta \lambda = 0.75$, $\Omega = \pi$, $\omega_c = 2$, $\gamma = \kappa = 10^{-2}$, and $T = 1$.}
    \label{fig-physical_hermitian_OTOC_FD_model}
\end{figure}

In the Floquet Dicke (FD) model, discussed in Sec. \ref{sec-floquet-dicke-model}, the system-cavity interaction strength is time-dependent, as given in Eq. (\ref{eq-Floquet_DickeHam}).  
We plot the variation of OTOC with time in Fig. \ref{fig-physical_hermitian_OTOC_FD_model}. 
It can be observed that the OTOC keeps on increasing to an exponential value for the time period considered. Upon fitting the curve with a logarithmic plot of the OTOC, the Lyapunov exponent comes out to be 0.325, which again follows the inequality of the Eq. (\ref{eq-lyapunov-inequality}). An interesting observation can be made when we compare this figure with the plot given in Fig. \ref{fig-Floquet_Dicke_model_ground_state_with_t}. The occasional dips in the value of the OTOC come at the points where the ground state energy of the FD model is constant, or the model is in a normal phase. At the points in time where the model shows superradiant phase transitions, the OTOC picks up and grows exponentially. 
This observation is in accordance with the behavior of OTOC observed in the previous models, where phase transition acted as the cause of the exponential rise of OTOC in the system.
Further, the FD model has correspondence with the $N$-qubit Dicke model if the system-cavity interaction strength is assumed to be a constant. Here, due to the choice of the parameters and time dependence of the system-cavity interaction strength, the FD model oscillates between the normal and the superradiant phases. However, the maximum value of the OTOC in the case of $N$-qubit Dicke model's superradiant phase is lesser than the same in the FD model's superradiant phase at longer times. This value keeps rising after every transition from the normal to the superradiant phase, which is not the case in the $N$-qubit Dicke model. 
The rapid oscillations between the normal and the superradiant phases could be the root cause of the rapid increment in the value of OTOC.
\subsection{Short-time, long-time, and intermediate-time behavior of OTOC}
\begin{figure}[!htb]
    \centering
    \includegraphics[width = 1\columnwidth]{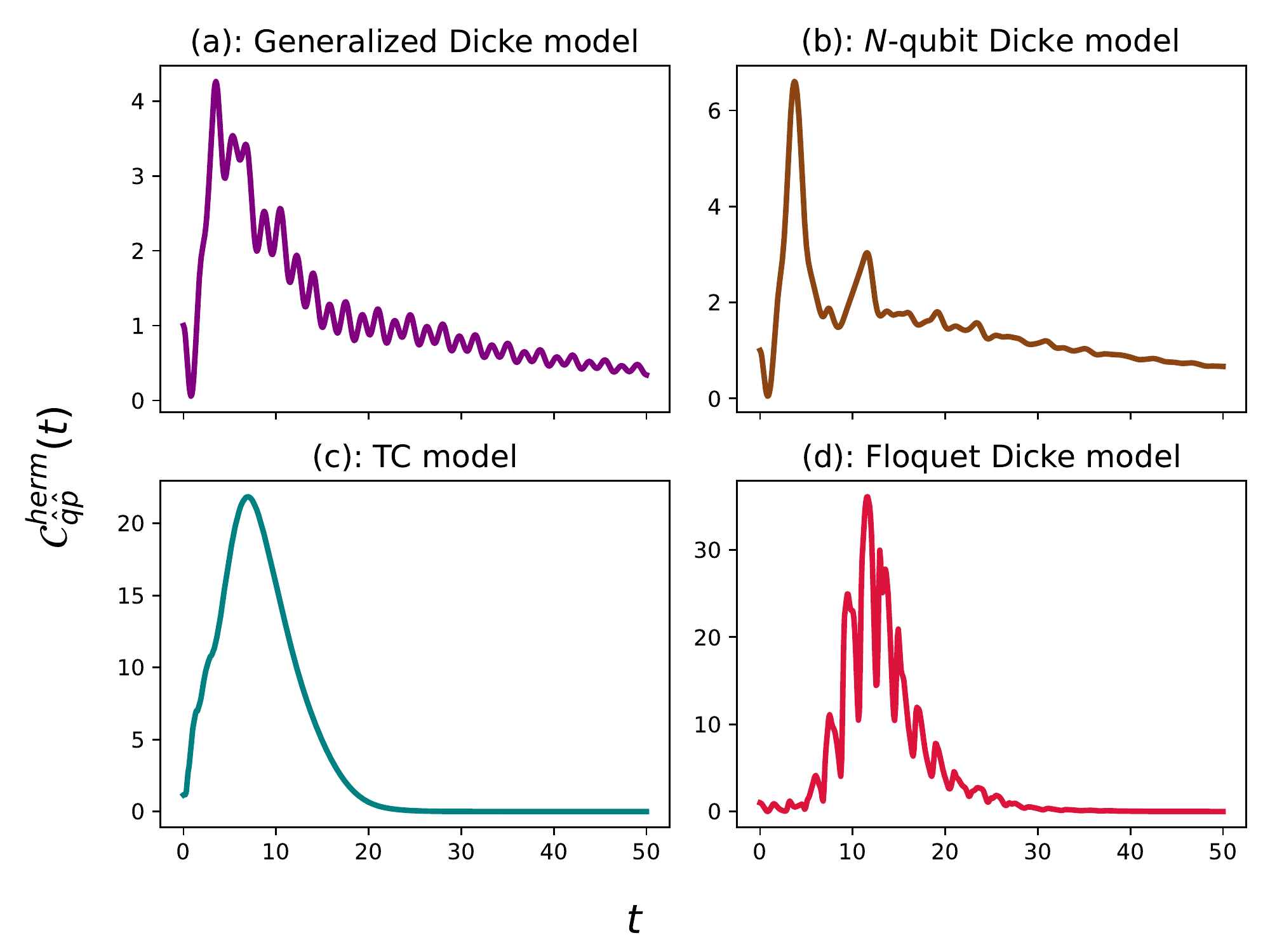}
    \caption{Variation of the OTOC $\mathcal{C}^{\text{herm}}_{\hat q \hat p} (t)$ with time in case of (a): generalized Dicke (with $\lambda = 1.5$, and $\lambda' = 1.2$), (b): $N$-qubit Dicke (with $\lambda = 1.5$), (c): TC model (with $\lambda = 4.5$), and (d): Floquet Dicke models (with $\lambda = 0.65$, $\Delta \lambda = 0.75$, and $\Omega = \pi$) at longer times. Here we have chosen $ N = 7, \omega_a = \omega_c = 2$, $\gamma = \kappa = 10^{-2}$, and $T = 1$. }
    \label{fig-Long_time_behavior_OTOC_TC_Model}
\end{figure}

We have calculated the OTOC of the various quantum optical models above. We now observe the variation of the OTOC in various time regimes together with the effect of dissipation at longer times. In Fig. \ref{fig-Long_time_behavior_OTOC_TC_Model}, we have plotted the long-time behavior of the OTOC. 
It can be observed that due to the dissipation in the evolution of the system with time, the OTOC decays after reaching an exponential value. Furthermore, the OTOC in the case of the Floquet Dicke model reaches the maximum value in comparison to the other models. The initial time up to which the OTOCs grow exponentially depicts the short-time behavior of the OTOCs, and these attain a global maximum value at the scrambling time $t^*$. In the case of the TC model, the scrambling time is around 9. The exponential growth of the OTOCs stops at the scrambling time. A scrambled system corresponds to a maximal entanglement entropy of any subsystem when the system is prepared in a pure state. The time regime in which the OTOC transitions from exponential growth to decay, around the scrambling time, represents the intermediate time regime. However, the intermediate time regime for the models considered here can not be explicitly pointed out because the transition from exponential growth to decay is very quick. In the long-time regime, we observe that due to dissipation in the dynamics of the system, the OTOCs decay. At longer times, the value of the OTOC for the $N$-qubit and generalized Dicke models is higher than the value of OTOCs for the Floquet Dicke and TC models, which are approximately zero. 

The long-time behavior of the OTOC is important to identify the chaotic dynamics of a system. The short-time exponential growth as an identifier of quantum chaos is not universal~\cite{ignacio} as it detects the instabilities in the system as well, and it may not be used as an absolute measure of the quantum chaos~\cite{Stransky_2023, Hummel_2019, Hirsch_2020_cameo}. This can be the reason that the TC model, which is an integrable model, shows an initial exponential growth of the OTOC. It was shown in~\cite{Stransky_2023} that the long-time behavior of the OTOC for a regular model is oscillatory, whereas, in the case of a chaotic system, the OTOC saturates to a fixed value with very small asymptotic oscillations. In this context, for the models studied here, we observe that the OTOC for the TC model drops quickly to zero due to dissipation. However, for the generalized Dicke and the $N$-qubit Dicke models, the OTOC drops after initial exponential growth but does not go quickly to zero; instead, it tries to saturate to a non-zero value with small oscillations.

\section{OTOC in the $N$-qubit Dicke model for all angular momentum sectors}
The Dicke model is shown to exhibit both the excited-state quantum phase transition and the thermal phase transition~\cite{Relano_TPT, Hirsch_JSTAT}. This depends on the angular momentum sectors of the Dicke model. When the maximum angular momentum sector is taken into account, that is, $j = N/2$, the excited-state quantum phase transition is observed, and a thermal phase transition occurs when all the sectors of the angular momentum are used. Consider the model given in Eq. (\ref{eq-gen_dicke_Ham_all}) for $\lambda = \lambda'$, at $T\to 0$. The model exhibits quantum phase transition at $\lambda = \lambda_c = \frac{\sqrt{\omega_a\omega_c}}{2}$. However, for finite temperatures, there exists a thermal phase transition. The critical temperature at which this transition occurs is accounted for using the following equation~\cite{Relano_TPT}
\begin{equation}
    \beta_c = \frac{2}{\omega_a}\tanh^{-1}\left(\frac{\omega_a\omega_c}{4\lambda^2}\right),
\end{equation}
where $\beta_c = 1/k_BT_c$. Incidentally, the value of critical temperature is one if, for example, we consider the values of $\omega_a = \omega_c = 2$ and $\lambda = 1.15$. This value of $\lambda$ is close to the value of $\lambda_c = 1.0$ for the $T\to 0$ case. 

An interesting thing to do would be to study the behavior of OTOC for the $N$-qubit Dicke model without restricting to the $j = N/2$ maximally symmetric angular momentum sector. This would entail taking into account all the sectors of the angular momentum, which in turn makes the computation very cumbersome.
\begin{figure}
    \centering
    \includegraphics[width = 1\columnwidth]{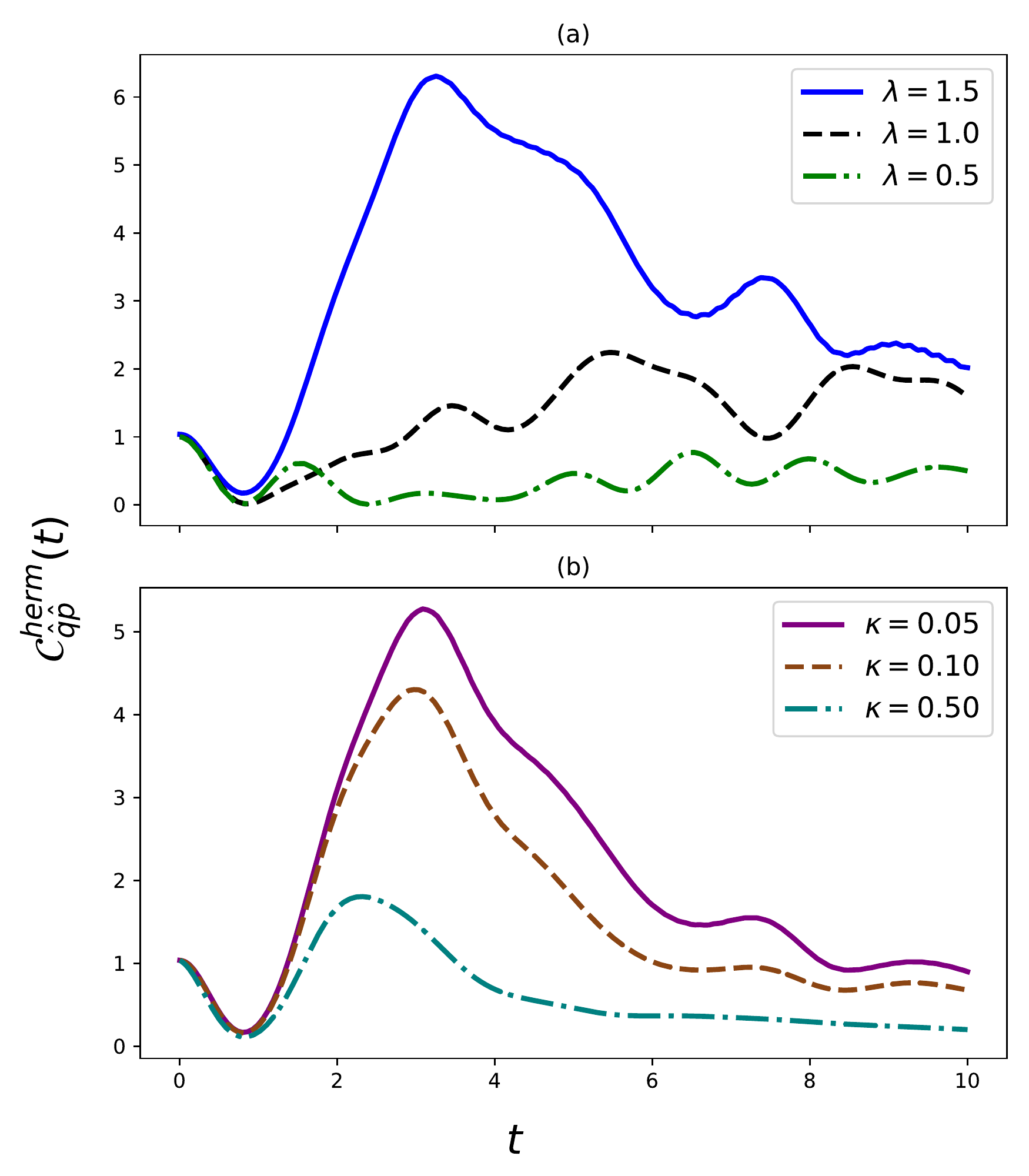}
    \caption{Variation of the OTOC with time for the $N$-qubit Dicke model considering all the angular momentum sectors. In subplot (a), we plot the variation of OTOC for different values of interaction strength (and for $\kappa = 10^{-2}$), and in (b), we plot the variation of OTOC for different cavity decay rates $\kappa$, keeping interaction strength $\lambda = 1.5$. The parameters chosen are:  $N=4$, $\omega_a = \omega_c = 2$, $\gamma = 10^{-2}$, and $T = 1$.}
    \label{fig:hermitian_otoc_full_Dicke_model}
\end{figure}
The corresponding analysis is depicted in Fig. \ref{fig:hermitian_otoc_full_Dicke_model}. 
We observe that the behavior of OTOC is similar to that observed in the case of the $N$-qubit Dicke model in Sec. III B for the collective angular momentum operator. In the regime where the thermal phase transition has occurred, corresponding to $\lambda \ge 1.15$, the OTOC shows an initial exponential behavior, whereas, below this critical atom-cavity interaction strength, the exponential behavior is suppressed (as in Fig. \ref{fig:hermitian_otoc_full_Dicke_model}(a)). The initial exponential behavior further diminishes for lower interaction strengths. We further observe the effect of dissipation on the variation of OTOC for fixed interaction strength, where the OTOC decays as we increase the cavity decay rate $\kappa$, as depicted in Fig. \ref{fig:hermitian_otoc_full_Dicke_model}(b).  

\section{Connection of OTOC with the characterizers of quantum optics}\label{sec-otoc_g2}
Here, we discuss the connection between the OTOC and the quantities characterizing quantum optics, primarily the second-order coherence function $g^{(2)}(t)$. Based on various conditions on this function, the Poissonian, sub-Poissonian, bunching, and anti-bunching effects of the cavity field can be characterized. The second-order coherence $g^{(2)}(t)$ function is one of the most important characterizers of a light source. It is defined as 
\begin{equation}
    \label{eq-g2_func}
    g^{(2)}(t) = \frac{\langle \hat a^\dagger (t') \hat a^\dagger(t) \hat a(t) \hat a(t')\rangle}{\langle\hat a^\dagger (t) \hat a(t)\rangle^2},
\end{equation}
where $\hat a$ and $\hat a^\dagger$ are the bosonic annihilation and creation operators, respectively, and $t = t' + \tau$. 
To this effect, in this section, we try to elaborate on the behavior of OTOC around these effects. 
Up to this point, we computed the OTOC using the operators $\hat q$ and $\hat p$ as given in Eq. (\ref{eq-adjoint_master_eq_qt}). We now make use of the annihilation operator $\hat a$  as the operators $A$ and $B$ given in Eq. (\ref{eq-otoc1}). The OTOC in this operator form can now be rewritten as 
\begin{equation}
    \label{eq-a_otoc}
    \mathcal{C}_{\hat a \hat a}(t, t') = \braket{[\hat a_t, \hat a_{t'}]^\dagger[\hat a_t, \hat a_{t'}]},
\end{equation}
where the the annihilation operator $\hat a$, is evolved to times $t'$ and $t = t'+\tau$, and then their commutator is calculated. Using Eq. (\ref{eq-adjoint-mastereq1}), the time evolution of $\hat a$ is given by the master equation 
\begin{equation}
    \label{eq-adjoint_master_eq_at_atp}
    \frac{d\hat a_t}{dt} = \mathcal{L}^\dagger(\hat a_t),
\end{equation}
where $\mathcal{L}^\dagger$ is same as given in Eq. (\ref{eq-adjoint_master_eq_qt}), where Lindblad jump operators are $L_1 = J_-$ and $L_2 = \hat a$ with $\nu_1 = \gamma$ and $\nu_2 = \kappa$ as the spontaneous and cavity decay rates, respectively. We use the above master equation for the time evolution of the operators $\hat a$ and $\hat a^\dagger$ in the subsequent sections. We now move on to derive a direct relation between the OTOC defined in Eq. (\ref{eq-a_otoc}) and the second-order coherence $g^{(2)}(t)$ function. 
\subsection{OTOC and the second-order coherence $g^{(2)} (t)$ function with superradiant phase transition}
To bring out an exact relation between the out-of-time-ordered-correlator function given in Eq. (\ref{eq-otoc1}) and the $g^{(2)}(t)$ function, we first rewrite the OTOC in Eq. (\ref{eq-a_otoc}) in its time-ordered and out-of-time-ordered components. This is similar to Eq. (\ref{eq-otoc3}) and can be given as 
\begin{equation}
    \label{eq-OTOC_g2}
    \mathcal{C}_{\hat a\hat a}(t, t') = \mathcal{D}_{\hat a\hat a}(t, t') + \mathcal{I}_{\hat a \hat a}(t, t') - 2\Re(\mathcal{F}_{\hat a \hat a}(t, t'),
\end{equation}
where for $t = t' + \tau$, we have 
\begin{align}
    \label{eq-DIF_g2_1}
    \mathcal{D}_{\hat a\hat a}(t, t') &= \langle \hat a^\dagger (t') \hat a^\dagger(t) \hat a(t) \hat a(t')\rangle, \nonumber \\
    \mathcal{I}_{\hat a\hat a}(t, t') &= \langle \hat a^\dagger (t) \hat a^\dagger(t') \hat a(t') \hat a(t)\rangle, ~\text{and} \nonumber \\
    \mathcal{F}_{\hat a \hat a}(t, t') &= \langle \hat a^\dagger (t) \hat a^\dagger(t') \hat a(t) \hat a(t')\rangle.
\end{align}
Upon a close inspection of the above equation and Eq. (\ref{eq-g2_func}), it can be seen that the time-ordered function $\mathcal{D}_{\hat a\hat a}(t, t')$ has a direct correspondence with the $g^{(2)}(t)$ function, and in fact, $\mathcal{D}_{\hat a\hat a}(t, t') = \langle\hat a^\dagger (t) \hat a(t)\rangle^2 g^{(2)}(t)$. Using this, a direct relation between the out-of-time-ordered correlator $\mathcal{C}_{\hat a\hat a}(t, t')$ and the second-order coherence $g^{(2)}(t)$ function can be written as 
\begin{equation}
    \label{eq-g2_otoc_connect1}
    \mathcal{C}_{\hat a\hat a}(t, t') = \langle\hat a^\dagger (t) \hat a(t)\rangle^2 g^{(2)}(t) + \mathcal{I}_{\hat a \hat a}(t, t') - 2\Re(\mathcal{F}_{\hat a \hat a}(t, t').
\end{equation}
The $g^{(2)}(t)$ function is a four-point correlation function of the field operators $\hat a$ and $\hat a^\dagger$, and in the model considered here, this function tells us the behavior of the light interacting with the atomic system. The OTOC characterizes the dynamics of the system; in particular, it measures how the evolution of an operator affects the dynamics of the system. 
Here, we are able to make a direct comparison between the $g^{(2)}(t)$ and the OTOC. 
Practically, the $g^{(2)}(t)$ can be measured numerically, and the correspondence between this function and the OTOC can pave the way to observe the chaotic properties of a quantum optical system experimentally. 
\begin{figure}[!htb]
    \centering
    \includegraphics[width = 1\columnwidth]{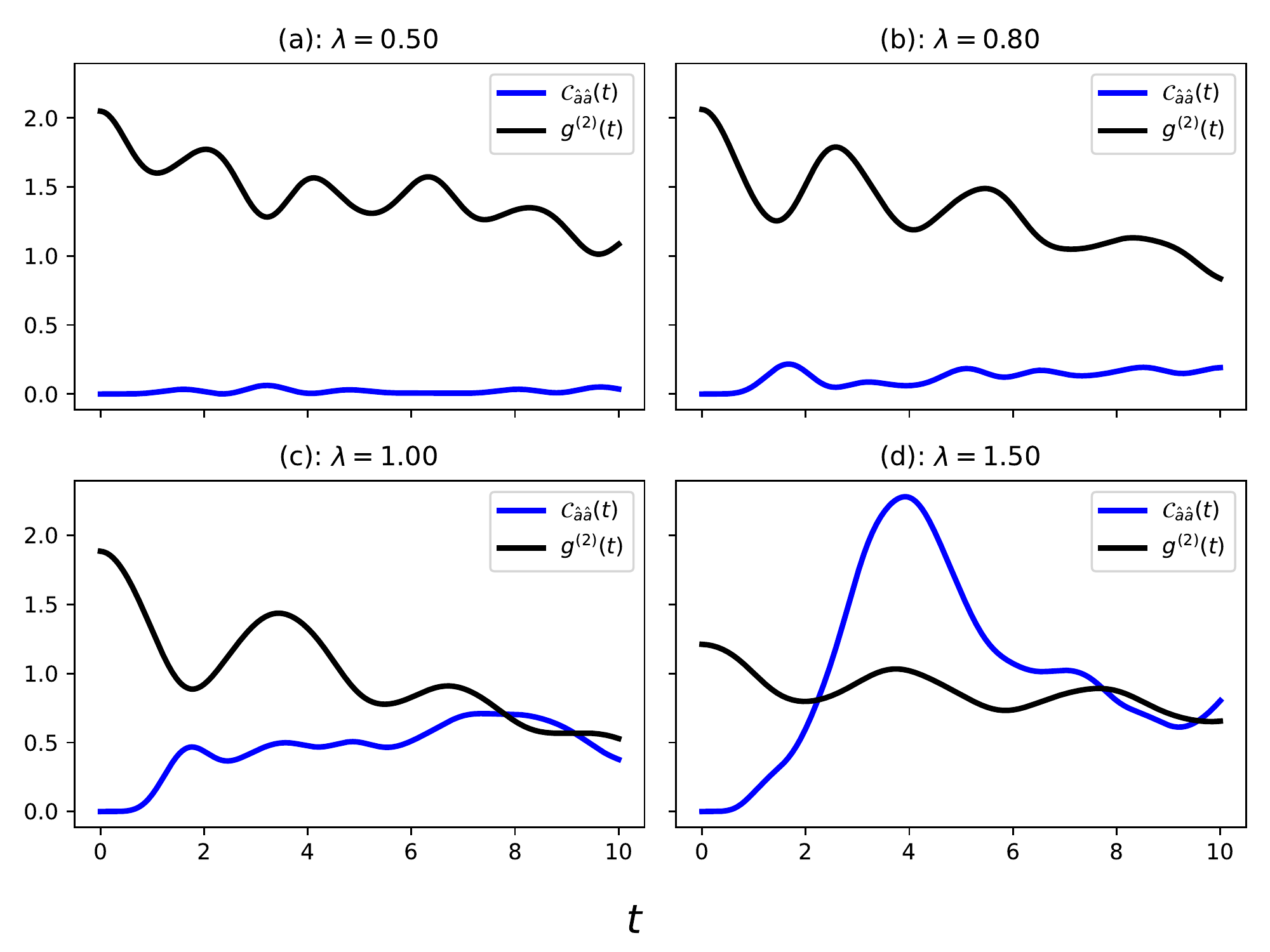}
    \caption{Variation of the OTOC $\mathcal{C}_{\hat a\hat a}(t)$ given in Eq. (\ref{eq-a_otoc}), and the second-order coherence $g^{(2)}(t)$ function (Eq. (\ref{eq-g2_func})) with time $t$ for $t' = 0$, for the $N$-qubit Dicke model. The parameters are chosen to be $N=7$, $\omega_a = \omega_c = 2$, $\gamma = \kappa = 10^{-2}$, and $T = 1$.}
    \label{fig-g2_func_superradiance_otoc}
\end{figure}%

Moreover, we try to re-establish the connection of the OTOC with the superradiant phase transition of the $N$-qubit Dicke model as given in Sec. \ref{sec-n_qubit_Dicke_otoc}. Here, the OTOC is calculated using the time evolution of the creation and annihilation operators $\hat a$ and $\hat a^\dagger$, respectively, instead of $\hat q$ and $\hat p$ operators used in Sec. \ref{sec-otoc_behavior}. For the choice of parameters here, the critical value of atom-field interaction strength $\lambda_c$ is one for superradiant phase transition. In Fig. \ref{fig-g2_func_superradiance_otoc}, it can again be observed that the OTOC has an exponential growth in the superradiant regime, and it is absent in the normal phase. 
An interesting comparison between the $g^{(2)}(t)$ function, OTOC, and the superradiant phase transition can be made in this case. The frequency of the oscillations in the variation of the $g^{(2)}(t)$ function and the initial value of this function decreases as the $N$-qubit Dicke model enters the superradiant phase. In the normal phase, the dips and the blips of the OTOC and $g^{(2)}(t)$ function do not match. However, in the superradiant phase, the fluctuations in the OTOC and $g^{(2)}(t)$ function are in tandem. 
Further, we explore the optical properties of light attached to the $g^{(2)}(t)$ function and analyze the behavior of OTOC in conjunction with these properties next. 

\subsection{Characteristics of light and the OTOC}
The second-order $g^{(2)}(t)$ coherence function can be used to characterize various properties of light. This includes bunching and anti-bunching of light, as well as super-Poissonian (or classical), Poissonian (or coherent), and sub-Poissonian (or non-classical) distributions of light. The conditions on the equal time second-order coherence $g^{(2)}(0)$ function, that is, whether $g^{(2)}(0)>1$, $g^{(2)}(0)=1$, or $g^{(2)}(0)<1$ identifies whether the photon distribution of the light is super-Poissonian, Poissonian, or sub-Poissonian, respectively. Furthermore, the conditions on the $g^{(2)}(t)$ function, that is, whether $g^{(2)}(t)<g^{(2)}(0)$, or $g^{(2)}(t)>g^{(2)}(0)$ determines the photon-counting statistics of light. In the time scale where $g^{(2)}(t)<g^{(2)}(0)$, the photons distribute themselves in bunches, known as the bunching effect of the light, whereas when $g^{(2)}(t)>g^{(2)}(0)$, there are fewer photons that can be detected in bunches; this is the anti-bunching effect of the light. 
\begin{figure}[!htb]
    \centering
    \includegraphics[width = 1\columnwidth]{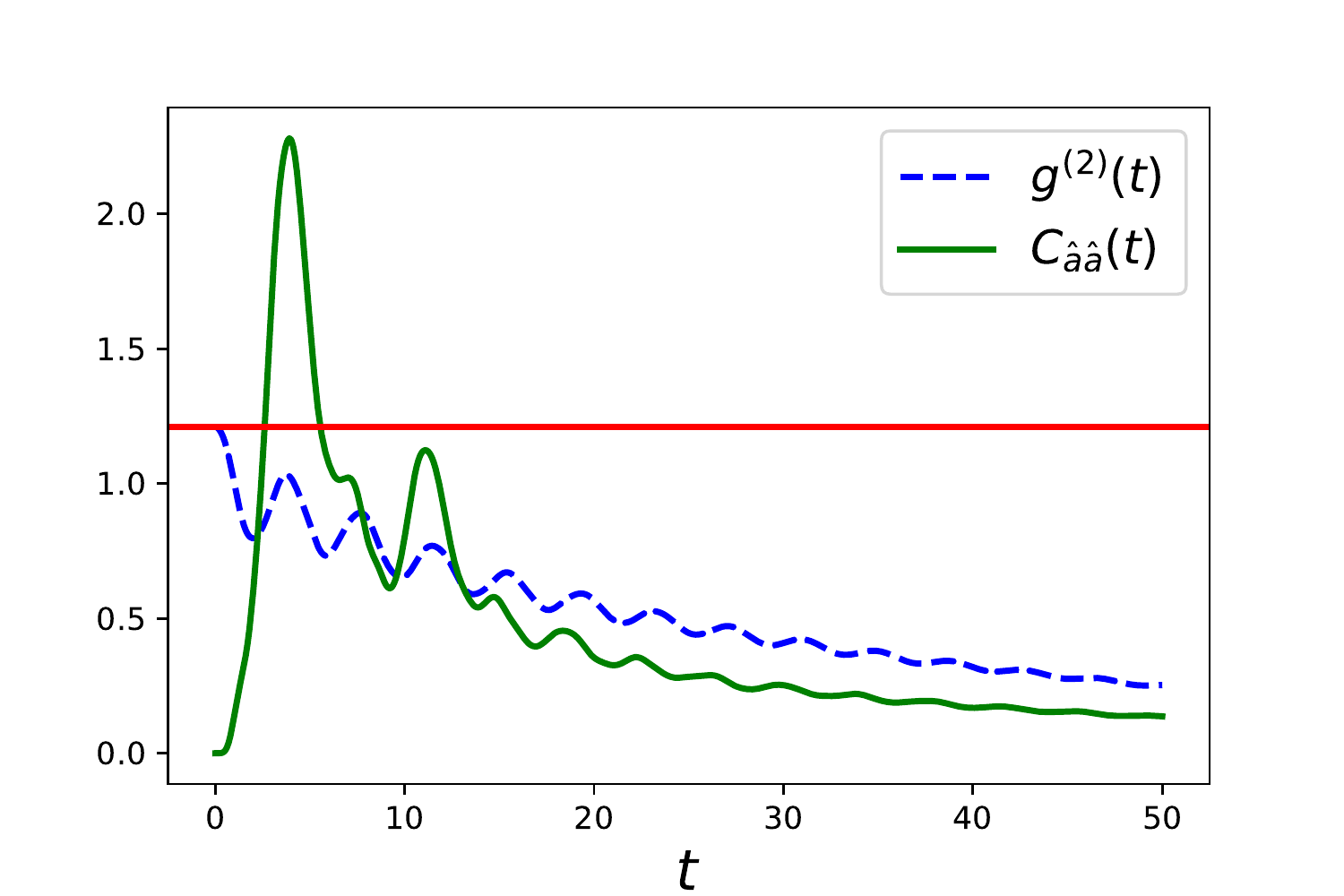}
    \caption{Variation of the OTOC $\mathcal{C}_{\hat a \hat a} (t)$, and the second-order coherence $g^{(2)}(t)$ function with time in case of the $N$-qubit Dicke model. The red line shows the value of equal time second-order coherence $g^{(2)}(0)$ function at $t' = 0$. Here we have chosen $N=7$, $\omega_a = \omega_c = 2$, $\lambda = 1.5$, $\gamma = \kappa = 10^{-2}$, and $T = 1$. }
    \label{fig-G2_func_with_otoc}
\end{figure}
In Fig. \ref{fig-G2_func_with_otoc}, the variation of OTOC together with the second-order coherence $g^{(2)}(t)$ function is shown with time for the $N$-qubit Dicke model. It can be observed that the initial time behavior of OTOC (from $t = 1$ to 3.5) in this case is exponential, and we get a value of $0.51$ for the Lyapunov exponent upon fitting the curve in this region. The light in the case considered has a super-Poissonian distribution of photons as $g^{(2)}(0)$ is greater than 1. Further, here at any time $t$, we have $g^{(2)}(t)<g^{(2)}(0)$. Therefore, the light is bunched in nature. This is expected as the state of the system is considered to be in the thermal state. Moreover, a connection between the $g^{(2)}(t)$ and OTOC $\mathcal{C}_{\hat a \hat a} (t)$ is observed. The changes of local maxima and minima in the plot of the $g^{(2)}(t)$ match with those of the OTOC. Further, in the long-time regime, both $g^{(2)}(t)$ function and OTOC saturates to a fixed value.

\section{Conclusion}
We have studied the quantum chaotic properties of a class of quantum optical models, {\it viz.}, the Dicke model, and its variants. To this end, we studied the generalized Dicke model, which is a multi-atom atom-cavity field interaction model. From this model, one can derive the two famous atom-cavity field interaction models, one using equal contributions from the co- and counter-rotating interaction terms, that is, the $N$-qubit Dicke model. The other one can be obtained using the rotating wave approximation, that is, the Tavis-Cummings model. Further, we studied the case of the time-dependent interaction strength between the atom and cavity field using the Floquet Dicke model. In all of the above models, we studied the boundaries of the phase transitions. In the case of the generalized, $N$-qubit, Tavis-Cummings, and Floquet Dicke models, we discussed the phase transition between the normal and the superradiant phases. We discussed the various forms of the out-of-time-ordered-correlators (OTOCs) present in the literature and the connection between them. This included the commutator and anti-commutator forms of OTOCs, regularized and physical OTOCs, and the relationship between them using Wigner-Yanase skew information. Further, we explored the connection between the out-of-time-ordered and the time-ordered correlators. 

We considered a non-unitary evolution of the operators in time to calculate OTOC using the adjoint form of the GKSL quantum master equation. The motivation for this was to bring out a more realistic picture of evolution, which is affected by factors like spontaneous emission and cavity decay. The OTOC for the class of quantum optical models discussed in this paper was calculated. We benchmarked the relationship between the various forms of OTOCs using quantum optical models. The Lyapunov exponent for the exponential behavior of OTOC was calculated, and the bound on the Lyapunov exponents was verified. 
It was observed that the exponential behavior of OTOC kicks in after the quantum phase transition of the models. The effect of dissipation could also be seen in the growth of OTOC; as we increase the dissipative factor, the OTOC decays. The long-time behavior of the OTOC was also explored. For the TC and the Floquet Dicke models, the OTOC decayed approximately to zero after the initial exponential rise because of dissipation. However, for the generalized and $N$-qubit Dicke models, the OTOC exhibited a tendency to saturate after an initial exponential rise. Taken in conjunction with the short-time exponential behavior of OTOC, this would suggest quantum chaotic behavior in the generalized and $N$-qubit Dicke models.
We also checked the variation of OTOC for the disordered Dicke model and its variant, the Richardson-Gaudin model. In these models, an exponential rise in the growth of OTOC was not observed.

We also established a relationship between the OTOC and the second-order coherence ($g^{(2)}(t)$) function using the time evolution of bosonic creation and annihilation operators through the adjoint GKSL master equation. The relationship between the $g^{(2)}(t)$ function and OTOC is important from an experimental point of view. In this case, we also studied the behavior of the OTOC with quantum phase transition of the $N$-qubit Dicke model together with the changes in the $g^{(2)}(t)$ function. It was observed that the frequency of the oscillations and the initial value of $g^{(2)}(t)$ function decreases as we increase the atom-cavity field interaction strength. Below the critical point of phase transition, where the Dicke model is in the normal phase, the dynamics of OTOC exhibit non-exponential behavior and becomes exponential in the superradiant phase. Further, the peaks and dips of the OTOC match with those of the $g^{(2)}(t)$ function in the superradiant region, which is not the case in the normal phase. Moreover, for the case considered in this paper, we observed that the behavior of light is super-Poissonian, and its photons are bunched in nature.

\section*{Acknowledgements}
S.B. acknowledges support from the Interdisciplinary Cyber-Physical Systems (ICPS) programme of the Department of Science and Technology (DST), India, Grant No.: DST/ICPS/QuST/Theme-1/2019/6 and DST/ICPS/QuST/Theme-1/2019/13. 

\bibliographystyle{apsrev4-1}
\bibliography{reference}

\end{document}